\documentclass[preprint,aps,showpacs,showkeys]{revtex4}
\usepackage{epsfig,amsmath,amssymb}
\begin{document}

\title{Universal properties of highly frustrated quantum magnets
       \\
       in strong magnetic fields}
\author{Oleg Derzhko$^{1,2,3}$,
        Johannes Richter$^{3,2}$,
        Andreas Honecker$^4$,
        and
        Heinz-J\"urgen Schmidt$^5$}
\affiliation{$^1$Institute for Condensed Matter Physics,
             National Academy of Sciences of Ukraine,
             1 Svientsitskii Street, L'viv-11, 79011, Ukraine\\
             $^2$Max-Planck-Institut f\"ur Physik komplexer Systeme,
             N\"othnitzer Stra{\ss}e 38, 01187 Dresden, Germany\\
             $^3$Institut f\"ur Theoretische Physik,
             Universit\"at Magdeburg,
             P.O.\ Box 4120, 39016 Magdeburg, Germany\\
             $^4$Institut f\"ur Theoretische Physik,
             Georg-August-Universit\"{a}t G\"ottingen,
             Friedrich-Hund-Platz 1, 37077 G\"ottingen, Germany\\
             $^5$Fachbereich Physik,
             Universit\"at Osnabr\"uck,
             Barbarastr.\ 7, 49069 Osnabr\"uck, Germany}

\date{December 12, 2006; revised February 13, 2007}

\begin{abstract}
The purpose of the present paper is two-fold. On the one hand,
we review some recent studies on the low-temperature
strong-field thermodynamic properties of frustrated quantum spin
antiferromagnets which admit the so-called localized-magnon eigenstates.
One the other hand, we provide some complementary new results.
We focus on the linear independence of the localized-magnon states, the
estimation of their degeneracy with the help of auxiliary classical
lattice-gas models and the analysis of the contribution of these states to
thermodynamics.
\end{abstract}



\maketitle

\clearpage

\section{Introduction}

Antiferromagnetically interacting quantum Heisenberg spins on regular
geometrically frustrated lattices have attracted much interest over the
past few decades \cite{rev1,rev2,rev_rich_etal}. On the one hand,
there has been
tremendous recent progress in synthesizing
corresponding magnetic materials \cite{lemmens}. On the other hand,
geometrically frustrated quantum antiferromagnets are
an excellent play-ground for studying novel quantum many-body
phenomena. We mention
quantum spin-liquid phases, order-by-disorder
phenomena, lattice instabilities to name just a few.
Application of an external magnetic field to a frustrated quantum
Heisenberg antiferromagnet introduces a new competition between
interactions in the spin system that may lead to further interesting
phenomena.  As an example we mention the half-magnetization plateau
stabilized by structural distortion in the pyrochlore lattice
\cite{penc,balents}.

The theoretical investigation of frustration effects in quantum spin
antiferromagnets usually meets new difficulties; e.g., the quantum Monte
Carlo method suffers from the sign problem for frustrated systems.
However, it is amazing that just owing to geometrical frustration some
possibilities for rigorous analysis emerge.  Recently, it has been
recognized that many geometrically frustrated lattices (including the
kagom\'{e} lattice, the checkerboard lattice and the pyrochlore lattice)
admit a simple class of exact eigenstates christened localized magnons
\cite{lm_01,lm_02}. These states become the ground states in strong
magnetic fields and they are relevant for many physical properties of a
wide class of highly frustrated quantum antiferromagnets in the
low-temperature strong-field regime
\cite{lm_01,lm_02,lm_03,lm_04,lm_05,lm_06,lm_07,lm_08,lm_09,lm_10,lm_11,lm_12,lm_13,lm_14,lm_15,lm_16,lm_17}.
In particular, the localized-magnon states are responsible for
magnetization jumps which the ground-state magnetization curve exhibits at
the saturation field \cite{lm_01,lm_02,lm_04,lm_09,lm_10,lm_11,lm_17}, may
lead to a high-field spin-Peierls lattice instability
\cite{lm_05,lm_12,lm_13}, and imply a residual ground-state entropy at the
saturation field \cite{lm_06,lm_07,lm_08}. Moreover, these states dominate
the low-temperature thermodynamics in the vicinity of the saturation field
\cite{lm_06,lm_07,lm_08,lm_14,rev_zhito,lm_16} and may lead to an
order-disorder phase transition of purely geometrical origin
\cite{lm_07,rev_zhito,lm_16}.

This paper focusses on new developments concerning
localized-magnon effects.  We mention several reviews on this subject
\cite{rev_rich_etal,rev_zhito,rev_rich} which, however, do not cover some
recent studies on the universal thermodynamic behavior which emerges at
low temperatures around the saturation field.
We will therefore concentrate
on the results obtained during the last two years
\cite{lm_14,lm_15,lm_16}. In passing, we will provide several
complementary  new results.

Although the effects of the localized magnons have not been observed
experimentally so far, recent studies on the spin-$1/2$
diamond-chain compound azurite
Cu$_{\rm{3}}$(CO$_{\rm{3}}$)$_{\rm{2}}$(OH)$_{\rm{2}}$
\cite{kikuchi_old,kikuchi} and the frustrated quasi-two-dimensional
spin-$1/2$ antiferromagnet Cs$_{\rm{2}}$CuCl$_{\rm{4}}$ \cite{radu}
represent closely related physics.

The present paper is organized as follows. In Section \ref{sec:Loc}, we
recall some basic results concerning the localized-magnon states which are
then used throughout the paper. Further, in Section \ref{indep}, we
discuss the linear independence of the localized-magnon states with the
smallest localization area \cite{lm_15} and their completeness as a basis
of the high-field ground states.
Next, in Section \ref{sec:LT}, we discuss the universal
thermodynamic behavior which the considered frustrated quantum
antiferromagnets exhibit at low temperatures in the vicinity of the
saturation field \cite{lm_14,lm_16}. We distinguish different types of the
universal behavior depending on the specific classical lattice-gas model
which represents the low-energy degrees of freedom of the spin system. We
focus on several representative models, namely, the diamond chain
(hard-monomer universality class), the frustrated two-leg ladder and the
kagom\'{e}-like chain (one-dimensional hard-dimer universality class), and
the frustrated bilayer lattice (two-dimensional hard-square universality class).
Finally, in Section \ref{sec:Concl}, we summarize and briefly comment on
some possibilities of experimental observation of the localized-magnon
effects.

\section{Localized-magnon states}

\label{sec:Loc}

\begin{figure}
\begin{center}
\includegraphics[clip=on,width=5.4cm,angle=0]{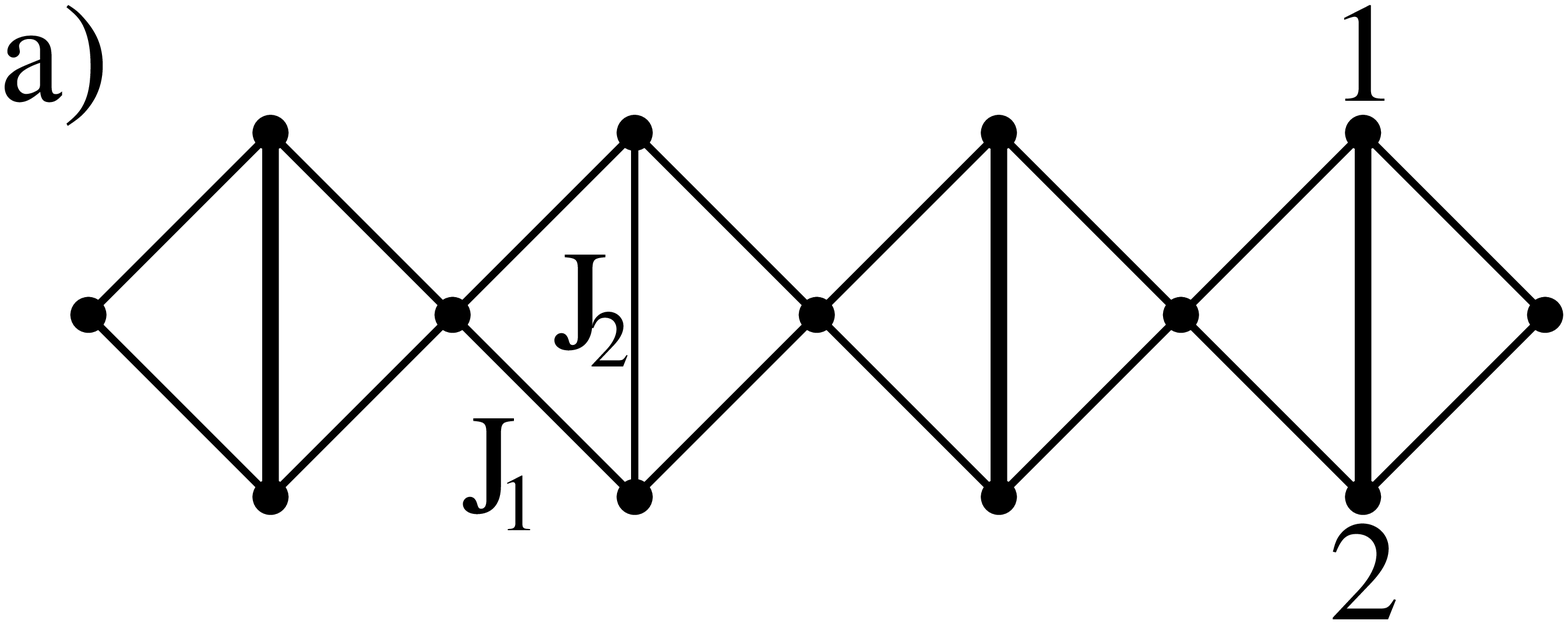}\\
\vspace{20mm}
\includegraphics[clip=on,width=9.50cm,angle=0]{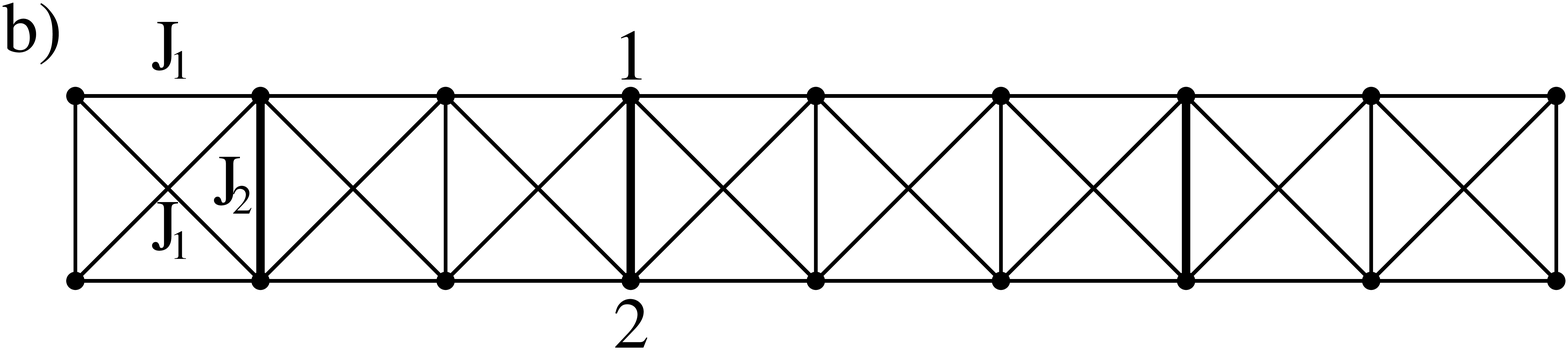}\\
\vspace{20mm}
\includegraphics[clip=on,width=9.50cm,angle=0]{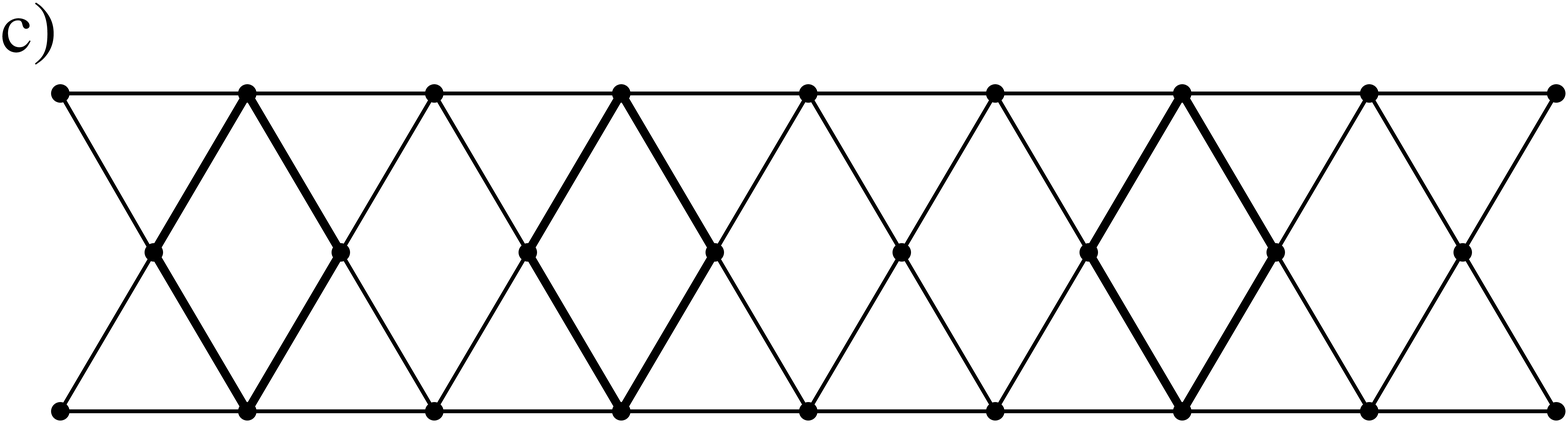}
\caption
{Three examples of one-dimensional spin lattices admitting
localized magnons:
(a) the diamond chain
(hard-monomer universality class),
(b) the frustrated two-leg ladder
(one-dimensional hard-dimer universality class),
(c) the kagom\'{e}-like chain, originally introduced in \cite{wksme00}
(one-dimensional hard-dimer universality class).}
\label{fig01}
\end{center}
\end{figure}

\begin{figure}
\begin{center}
\includegraphics[clip=on,width=8.0cm,angle=0]{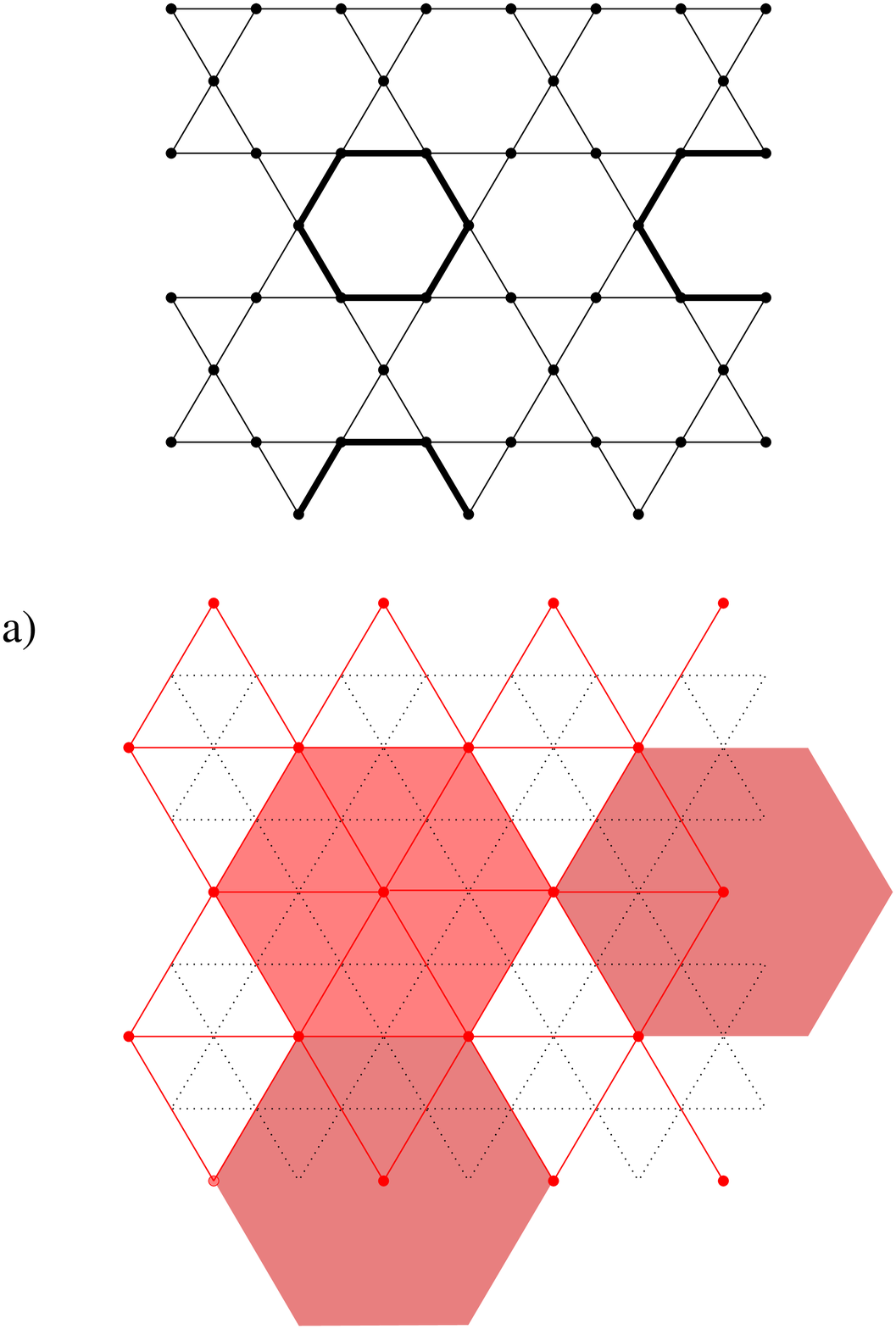}
\includegraphics[clip=on,width=8.0cm,angle=0]{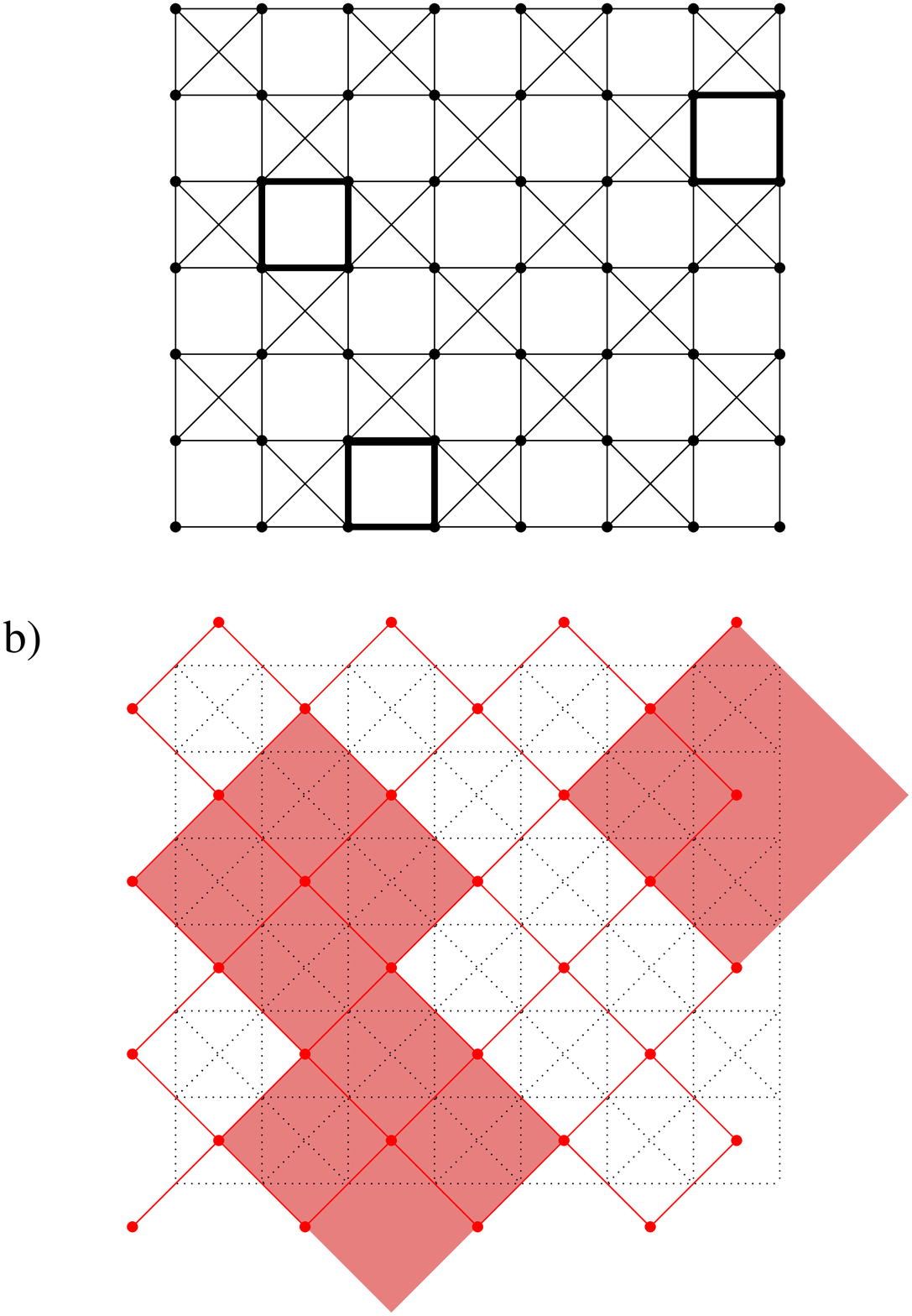}\\
\vspace{15mm}
\includegraphics[clip=on,width=10.0cm,angle=0]{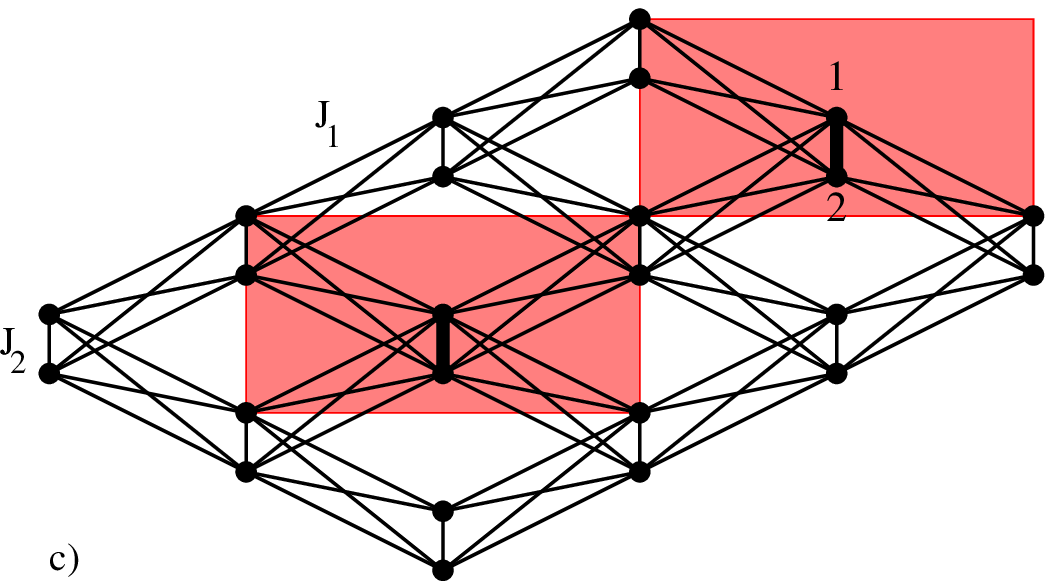}
\caption
{Three examples of two-dimensional spin lattices admitting
localized magnons:
(a) the kagom\'{e} lattice
(hard-hexagon universality class),
(b) the checkerboard lattice
(large-hard-square universality class),
(c) the frustrated bilayer lattice
(hard-square universality class).
We also show auxiliary lattice-gas models
(hard hexagons on a triangular lattice,
large hard squares on a square lattice
and
hard squares on a square lattice)
which describe low-energy degrees of freedom
of the spin models in strong magnetic fields.}
\label{fig02}
\end{center}
\end{figure}

In this paper we study the Heisenberg model
on several geometrically frustrated lattices.
Some of these lattices are shown in Figs.~\ref{fig01},~\ref{fig02}.
The Heisenberg Hamiltonian of $N$ quantum spins of length $s$ reads
\begin{eqnarray}
\label{01}
H=
\sum_{(nm)}J_{nm}\left(\frac{1}{2}\left(s_n^+s_m^-+s_n^-s_m^+\right)
+\Delta s_n^zs_m^z\right)
-h \, S^z.
\end{eqnarray}
Here the sum runs over the bonds (edges)
which connect the neighboring sites (vertices)
on the spin lattice (see Figs.~\ref{fig01},~\ref{fig02}),
$J_{nm}>0$ are the antiferromagnetic exchange integrals
between the sites $n$ and $m$,
$\Delta\ge 0$ is the exchange interaction anisotropy parameter, $h$ is the
external magnetic field, and
$S^z=\sum_ns_n^z$ is the $z$-component of the total spin (magnetization).
Note that the operator $S^z$ commutes with the Hamiltonian (\ref{01}).
For the examples shown in Figs.~\ref{fig01}a, \ref{fig01}b, and \ref{fig02}c
the exchange integrals take two values,
namely, $J_2$ for the vertical bonds
and $J_1$ for all other bonds;
for the other lattices shown in Figs.~\ref{fig01},~\ref{fig02}
they are uniform $J_{nm}=J$.
The exchange integrals  $J_1$ and $J_2$ may also be assumed to
satisfy certain relations, see, e.g., Ref.~\cite{lm_14} and below.
Although the present analysis can be performed for arbitrary values of $s$
and $\Delta$, in what follows we concentrate on the  extreme quantum case
$s=1/2$ and isotropic interactions $\Delta=1$,
in particular while performing exact diagonalization for finite systems \cite{footnote1}.

From Refs.~\cite{lm_01,lm_02,rev_rich_etal,rev_rich} we know
that the Heisenberg antiferromagnet (\ref{01})
on the lattices shown in Figs.~\ref{fig01},~\ref{fig02},
as well as on some other lattices like
the dimer-plaquette chain,
the sawtooth chain,
another kagom\'{e}-like chain
\cite{lm_02,lm_14,ahllt98}
(one-dimensional systems),
the square-kagom\'{e} lattice,
the star lattice
(two-dimensional systems),
and the pyrochlore lattice
(three-dimensional system),
support localized-magnon eigenstates.
The magnon may be localized  on the (vertical) bond
as for the lattices shown in Figs.~\ref{fig01}a, \ref{fig01}b,
and \ref{fig02}c, on the ${\sf{V}}$-part of the sawtooth chain,
or on the even polygon (hexagon, square etc.)
from which other lattices are built
(kagom\'{e}, pyrochlore, checkerboard, square-kagom\'{e} etc.),
see Figs.~\ref{fig01}c, \ref{fig02}a, and \ref{fig02}b.
Then the explicit expression for the localized-magnon state reads
\begin{eqnarray}
\label{02}
\vert 1{\rm{lm}}\rangle
=\vert{\rm{lm}}\rangle_{l}\vert s,\ldots,s\rangle_{e},
\end{eqnarray}
where
$\vert{\rm{lm}}\rangle_{l}
=
(1/\sqrt{2})
(\vert s\rangle_1 \vert s-1\rangle_2
-
\vert s-1\rangle_1 \vert s\rangle_2)$
for the lattices in which the magnon is trapped on the vertical bond
or
$\vert{\rm{lm}}\rangle_{l}
\propto
\sum_{m=1}^L
(-1)^ms_m^-
\vert s\rangle_1 \ldots \vert s\rangle_L$
for the lattices in which the magnon is trapped on the $L$-site polygon
and
$\vert s,\ldots,s\rangle_{e}$
denotes the ferromagnetically polarized environment.
By direct calculation one can convince oneself
that the state (\ref{02}) is indeed an eigenstate of the Hamiltonian (\ref{01})
with the eigenvalue $E_{{\rm{FM}}}-\epsilon_1$
where
$E_{{\rm{FM}}}$ is the energy of ferromagnetically polarized lattice
and
$\epsilon_1^{\rm diamond}=s(J_2+\Delta(2J_1+J_2))$,
$\epsilon_1^{\rm ladder}=s(J_2+\Delta(4J_1+J_2))$,
$\epsilon_1^{\mbox{\scriptsize kagom\'e}}=2s(1+2\Delta)J$,
and
$\epsilon_1^{\rm bilayer}=s(J_2+\Delta(8J_1+J_2))$
for the diamond chain, the frustrated two-leg ladder,
the kagom\'{e}-like chain, and the frustrated bilayer lattice,
respectively (here we put $h=0$ in (\ref{01})).

Alternatively,
one may diagonalize the Hamiltonian (\ref{01})
in the one-magnon subspace with $S^z=Ns-1$
to find that one of the magnon excitation branches is flat (dispersionless).
For the diamond chain, the frustrated two-leg ladder,
the kagom\'{e}-like chain, and the frustrated bilayer lattice
the dispersionless one-magnon energy is given by
$\Lambda_k^{\rm diamond}=-s(J_2+\Delta(2J_1+J_2))+h$,
$\Lambda_k^{\rm ladder}=-s(J_2+\Delta(4J_1+J_2))+h$,
$\Lambda_k^{\mbox{\scriptsize kagom\'e}}=-2s(1+2\Delta)J+h$,
and
$\Lambda_{\bf{k}}^{\rm bilayer}=-s(J_2+\Delta(8J_1+J_2))+h$,
respectively (note the correspondence of $\Lambda_k$ with $\epsilon_1$).
For some lattices
(the sawtooth chain, the kagom\'{e}-like chains, the kagom\'{e} lattice, the
checkerboard lattice etc.)
the dispersionless magnon band is the lowest one,
for other lattices it may become the lowest one
if certain relations on the antiferromagnetic exchange constants are imposed.
In particular, for the diamond chain and the frustrated two-leg ladder
we have to assume $J_2\ge 2J_1$,
whereas for the frustrated bilayer lattice
we have to assume $J_2\ge 4J_1$.
If equality in the imposed relations holds,
the dispersive higher-energy magnon band touches
the dispersionless lowest-energy band at some values of the wave-vector
(this also occurs for the kagom\'{e}-like chains, the kagom\'{e} lattice,
and the checkerboard lattice, but not for the sawtooth chain).

We pass to the subspaces with total $S^z=Ns-2,\ldots,Ns-n_{\max}$.
Here $n_{\max}$ is the number of the isolated localized magnons,
each occupying the smallest possible area, for the closest packing.
$n_{\max}$ depends on the lattice
and equals $N/3$, $N/4$, $N/6$, and $N/4$
for the diamond chain, the frustrated two-leg ladder,
the kagom\'{e}-like chain, and the frustrated bilayer lattice, respectively.
Evidently, we can construct eigenstates of the Hamiltonian (\ref{01})
of the form
\begin{eqnarray}
\label{03}
\vert 2{\rm{lm}}\rangle
=\vert{\rm{lm}}\rangle_{l_1}\vert{\rm{lm}}\rangle_{l_2}
\vert s,\ldots,s\rangle_{e},
\;\;\;
\ldots,
\;\;\;
\vert n_{\max}{\rm{lm}}\rangle
=\vert{\rm{lm}}\rangle_{l_1}\ldots\vert{\rm{lm}}\rangle_{l_{n_{\max}}}
\vert s,\ldots,s\rangle_{e},
\end{eqnarray}
which in the zero-field case $h=0$ have the energies
$E_{{\rm{FM}}}-2\epsilon_1, \;\ldots,\;
E_{{\rm{FM}}}-n_{\max}\epsilon_1$, respectively, provided
that the trapping cells of the localized magnons cannot be
directly connected.
Obviously, there are many other eigenstates in each of these
subspaces. The special importance of the localized-magnon states is
due to the fact that they may become ground states (or at least
low-lying states) in their respective subspaces. In Refs.\
\cite{lm_01,lm_03} it was proven under some quite general
assumptions that the localized-magnon states are indeed
lowest-energy states in the corresponding sectors of
$S^z=Ns,\ldots,Ns-n_{\max}$. More precisely, if we denote the
minimal energy within the subspace with $S^z=Ns-n$ by $E_{\min}(n)$,
the following inequality holds $E_{\min}(n)\ge
(1-n)E_{\rm{FM}}+nE_{\min}(1)
=E_{\rm{FM}}-n(E_{\rm{FM}}-E_{\min}(1)) =E_{\rm{FM}}-n\epsilon_1$
for all $n=0,1,\ldots,2Ns$ and
spin-$s$ Heisenberg systems with sufficiently general
coupling schemes \cite{lm_01,lm_03}. The energy of $n$
localized-magnon states is given by the expression on the r.h.s.\ of
this inequality (for $n=1,\ldots,n_{\max}$) and hence we conclude
that the localized magnons are lowest-energy states in the subspaces
with $S^z=Ns-1,\ldots,Ns-n_{\max}$.

\begin{figure}
\begin{center}
\includegraphics[clip=on,width=0.48\columnwidth,angle=0]{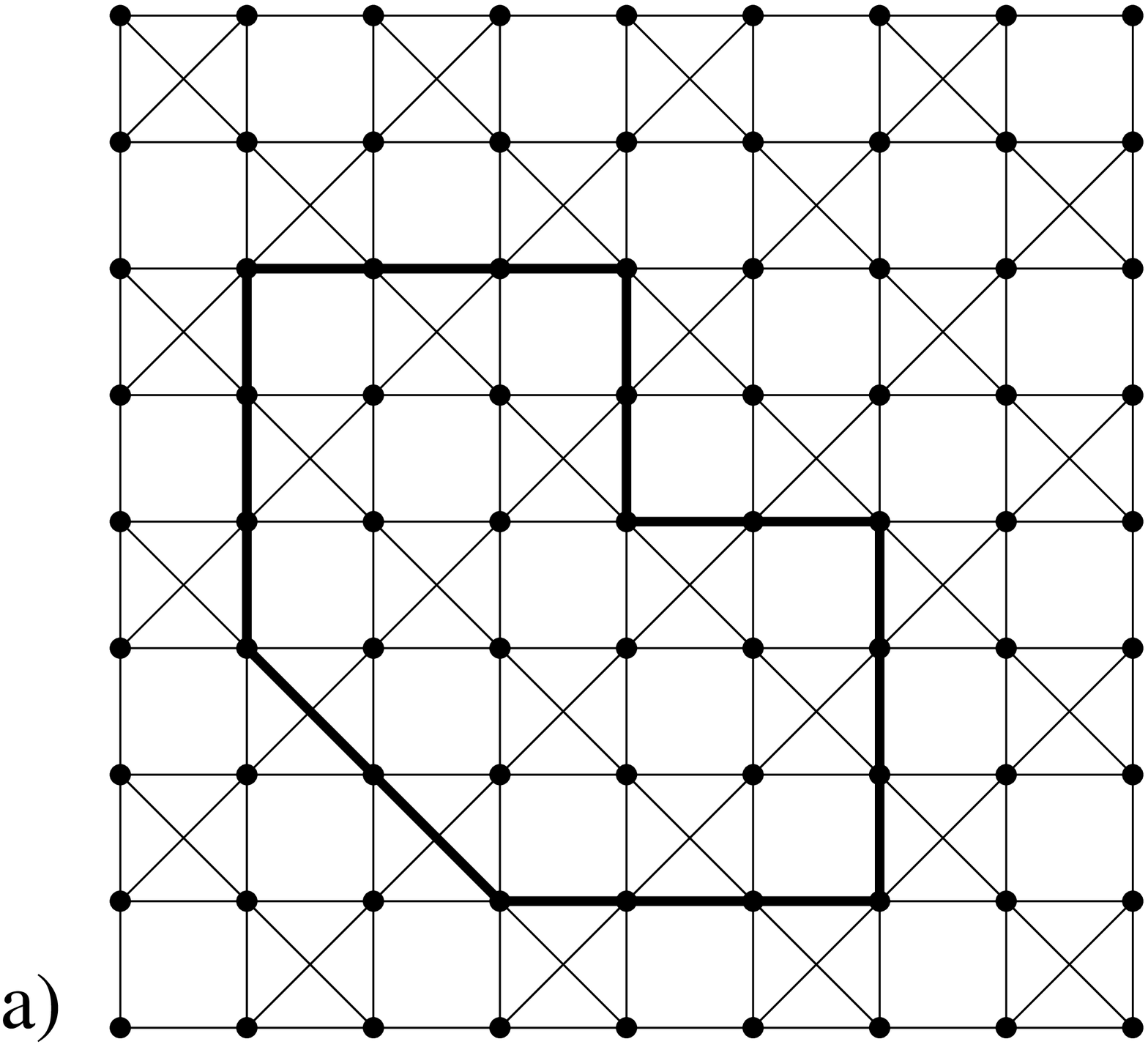}
\hfill
\includegraphics[clip=on,width=0.48\columnwidth,angle=0]{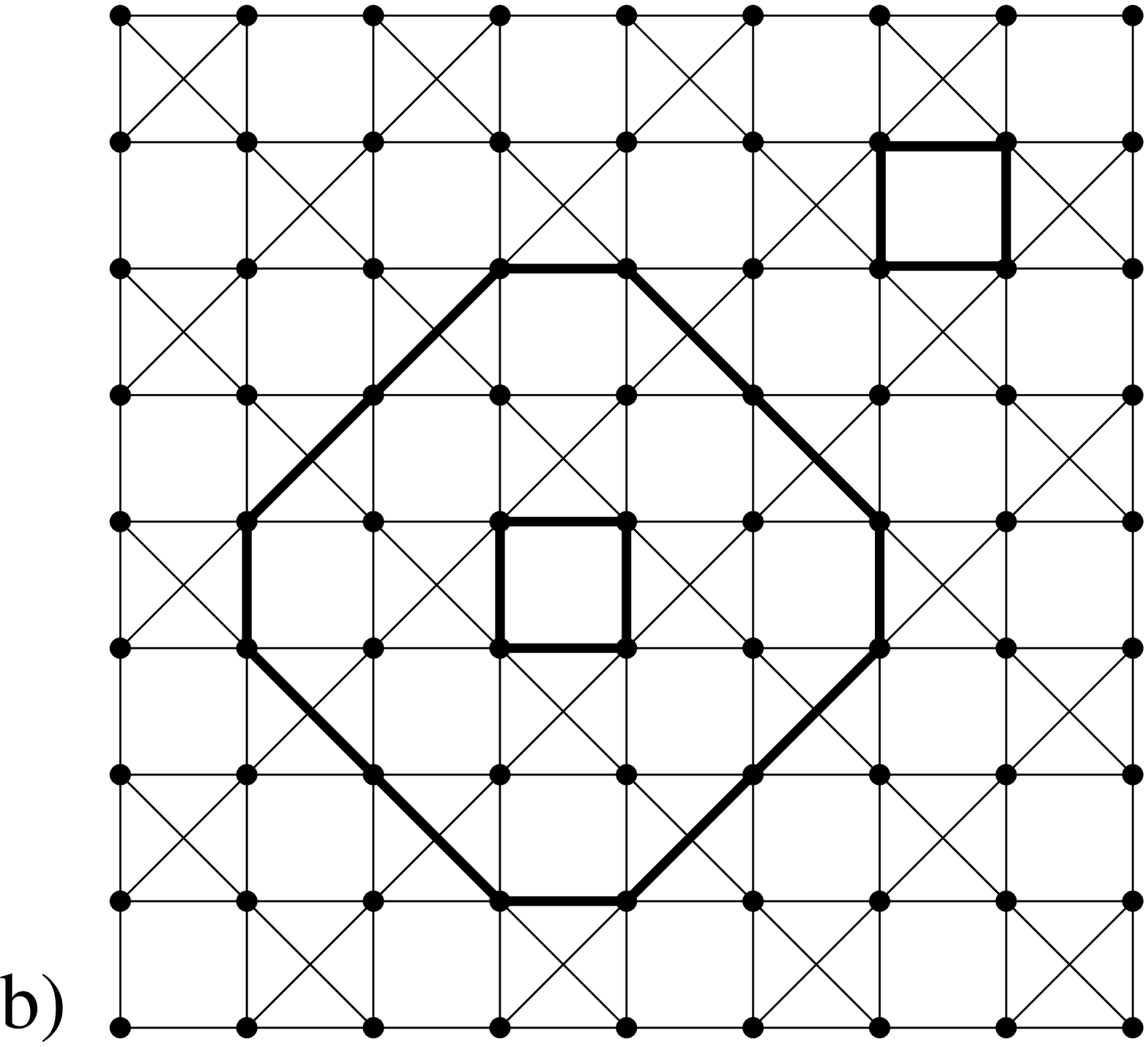}\\
\caption
{The checkerboard lattice.
(a) A localized magnon which occupies a larger area than the smallest
possible one.
(b) A three-magnon state which is not a large-hard-square state
because of the nested `defect' state at the lower left side of the lattice;
one magnon is localized on each of the three loops.}
\label{fig03}
\end{center}
\end{figure}

We note
that the localized-magnon states (\ref{02}), (\ref{03}) are highly degenerate states.
Obviously, the localized magnons of smallest area can be placed on a lattice in many ways.
Moreover,
for some lattices there are other states which have the same energy
as the localized magnons of smallest area. In the following discussion
we will concentrate on the checkerboard lattice and refer
to Refs.~\cite{lm_07,rev_zhito} for related remarks on the
kagom\'e lattice.
First of all we notice that for many lattices,
we can construct localized magnons occupying a larger area
in addition to the localized magnon of smallest area, see the
example in Fig.~\ref{fig03}a.
It is important to note that such eigenstates can be viewed as linear
combinations of the simpler eigenstates corresponding to the localized
magnons of smallest area. Moreover, in two and higher dimensions there may
be additional topological effects. In particular, for some lattices one
can construct nested objects (defect states),
compare the example
in Fig.~\ref{fig03}b.
Furthermore, for such lattices there may be additional localized magnons
with nontrivial winding if periodic boundary conditions are imposed
\cite{rev_zhito}. We shall come back to this issue when discussing the
degeneracy of the ground states in the subspaces
$S^z=Ns-2,\ldots,Ns-n_{\max}$ for finite systems using exact
diagonalization. In particular, we will see that the localized magnons of
smallest area linearly span the total ground state space in the sectors
$S^z=Ns-1,\ldots,Ns-n_{\max}$ for some lattices (e.g., the diamond chain
and the frustrated two-leg ladder with $J_2>2J_1$, the sawtooth chain, or
the frustrated bilayer lattice with $J_2>4J_1$).

One is thus led to the task of enumerating all many-localized-magnon
states. Let us denote the number of possibilities to put $n$ isolated
localized magnons, each occupying the smallest possible area, on a lattice
with $N$ sites by $g_N(n)\ge 1$. Furthermore note that the energy of a
state with $n$ localized magnons reads
\begin{eqnarray}
\label{04}
E_n(h)=E_{\rm{FM}}-h\,N\,s-n\,(\epsilon_1-h)
\end{eqnarray}
in the presence of an external magnetic field.
Thus, at the saturation field
$h=h_1=\epsilon_1$ we find that
the energy $E_n(h)$ does not depend on the number of localized magnons $n$.
Hence, the total ground-state degeneracy at $h=h_1$
is at least ${\cal{W}}=\sum_{n=0}^{n_{\max}}g_N(n)$
provided the considered localized-magnon states are linearly
independent (see Ref.~\cite{lm_15} and the next section).
As a result of this degeneracy one obtains the
following universal properties:
1)~the ground-state magnetization curve exhibits the jump at $h_1$
between the values $Ns-n_{\max}$ and $Ns$;
this jump is accompanied by a plateau at the foot of the jump;
2)~since ${\cal{W}}$ scales exponentially with $N$
the ground-state entropy per site at $h_1$
does not vanish
but remains finite, namely at least $k\ln{\cal{W}}/N>0$.
Moreover,
3)~the lattice is unstable with respect to a deformation,
which preserves the symmetry required for the existence of localized magnons
and lowers the magnetic energy linearly with respect to the displacement.
As a result, a field-tuned structural instability may take place.
We do not discuss these issues further;
more details can be found, e.g., in Refs.\ \cite{rev_rich_etal,rev_rich}.
Instead we turn to the localized-magnon effects
at finite but low temperatures
and strong magnetic fields around the saturation field.

\section{Linear independence and completeness}

\label{indep}

As discussed in the last section, the localized-magnon states constitute a highly degenerate
ground-state manifold at the saturation field $h=h_1$. One may expect that these states lead
to a dominant contribution to the low-temperature thermodynamics for magnetic fields  in the vicinity
of the saturation field.
The contribution of the localized-magnon states
to the partition function
can be written in the form
\begin{eqnarray}
\label{05}
Z_{{\rm{lm}}}(T,h,N)
=\sum_{n=0}^{n_{\max}}g_N(n)\exp\left(-\frac{E_n(h)}{kT}\right)
=\exp\left(-\frac{E_{{\rm{FM}}}-hNs}{kT}\right)
\sum_{n=0}^{n_{\max}}g_N(n)\exp\left(\frac{\mu}{kT}n\right),
\end{eqnarray}
where Eq.~(\ref{04}) was used
and $\mu=\epsilon_1-h=h_1-h$.
However, several questions,
briefly touched already in the previous section,
have to be discussed
before determining the degeneracy of the localized-magnon states $g_N(n)$
and analyzing the thermodynamic properties on the basis of Eq.~(\ref{05}).

First, we have to clarify
whether the set of localized magnons of smallest area
is linearly independent in each sector of $S^z=Ns,\ldots,Ns-n_{\max}$.
Only if this is the case,
{\em{all}} localized magnons of smallest area contribute
to the partition function of the spin system
and we do not have to take care about the states
which are their linear combinations
(like the state shown in Fig.~\ref{fig03}a).
Bearing in mind the thermodynamic limit $N\to\infty$
we need at least linear independence in all sectors of $S^z$
except probably in a few (finite number) sectors of $S^z$.

Although we do not have a general theory of linear independence or dependence of
the localized-magnon states of smallest area,
a detailed and systematic analysis 
for a wide class of lattices hosting localized magnons
was presented in Ref.~\cite{lm_15} including
rigorous proofs for lattices in one and two dimensions.
We discuss briefly these findings below following Ref.~\cite{lm_15}.
In this discussion we will refer to a localized magnon
of smallest area simply by ``localized magnon''.
It appears convenient to group the frustrated lattices into several classes
and to examine linear independence for each class separately.
Thus, we distinguish the following classes:
the orthogonal type
(the diamond chain (Fig.~\ref{fig01}a),
the dimer-plaquette chain,
the frustrated two-leg ladder (Fig.~\ref{fig01}b),
the square-kagom\'{e} lattice,
the frustrated bilayer lattice (Fig.~\ref{fig02}c));
the isolated type
(the sawtooth chain,
the kagom\'{e}-like chain I (Fig.~\ref{fig01}c)
and
the kagom\'{e}-like chain II);
the codimension 1 type
(the kagom\'{e} lattice (Fig.~\ref{fig02}a),
the star lattice,
the checkerboard lattice (Fig.~\ref{fig02}b));
and the higher codimension type (the pyrochlore lattice).

To examine the linear independence
of a finite sequence of $k$ vectors in some Hilbert space
it is useful to introduce the $k\times k$ Gram matrix
${\bf{G}}=\vert\vert G_{ij}\vert\vert$
of all scalar products.
The rank of ${\bf{G}}$ equals the dimension of the linear span of the
considered set, {\it i.e.}, the number of linearly independent states,
and the set is linearly independent iff ${\rm{det}}{\bf{G}}>0$.
We call the dimension of the null space of ${\bf{G}}$ the codimension
and hence the codimension will equal the number of independent linear relations
between $n$ localized-magnon states.

It can be proven (theorem 1 of Ref.~\cite{lm_15}) that
if the set of localized-magnon states is linearly independent
in the sector $S^z=Ns-1$,
then it is also linearly independent
in the sectors $S^z=Ns-n$, $n=2,\ldots,n_{\max}$.
Therefore, in many cases
it is sufficient to consider only localized one-magnon states.
In particular, for the lattices which belong to the orthogonal type,
the cells in which the localized magnons are concentrated are disjoint
and any two different localized one-magnon states are orthogonal.
As a result,
$G_{ij}\propto\delta_{ij}$ and according to the Gram criterion
all localized one-magnon states
(and hence all localized $n$-magnon states for every $n=1,\ldots,n_{\max}$)
are linearly independent.
Consider next the lattices which belong to the isolated type.
It can be proven (theorem 3 of Ref.~\cite{lm_15})
that if the trapping cell contains a site
which is not contained in any other trapping cell
(``isolated site'')
then all localized one-magnon states
(and hence all localized $n$-magnon states for every $n=1,\ldots,n_{\max}$)
are linearly independent.

We pass to the lattices of the codimension 1 type.
Suppose that every spin site is contained in exactly two different trapping cells
(this is just the case of the lattices of the codimension 1 type,
e.g., the kagom\'{e} lattice (Fig.~\ref{fig02}a),
the checkerboard lattice (Fig.~\ref{fig02}b) etc.).
Then it can be proven
that there is at most one linear relation between localized one-magnon states
(theorem 4 of Ref.~\cite{lm_15}).
Moreover, in such a case the set of localized-magnon states
in the subspaces $S^z=Ns-n$, $n=2,\ldots,n_{\max}$
is linearly independent (theorem 5 of Ref.~\cite{lm_15}).
Note that boundary conditions are crucial in this case.
As was pointed out already in Ref.~\cite{rev_zhito}, there is indeed
one relation between localized one-magnon states for the kagom\'{e} lattice
if periodic boundary conditions are imposed. For the checkerboard lattice with
periodic boundary conditions there is a similar 
linear relation \cite{lm_15}. By contrast, all localized
one-magnon states are linearly independent if one considers
the kagom\'e or checkerboard lattice with open boundary conditions.
Note that for the pyrochlore lattice 
there is more than one linear relation 
between one-magnon states localized on hexagons 
and the linear relations between such
localized-magnon states exist also in the subspaces $S^z=Ns-n$ for $n>1$ \cite{lm_15}.

The second important issue
concerns the completeness of
the localized-magnon states
in each sector of $S^z=Ns,\ldots,Ns-n_{\max}$.
More precisely, a quantitative description of the low-temperature
thermodynamics of the spin model in the vicinity of the saturation field
is obtained from the smallest-size localized magnons via Eq.~(\ref{05})
only if there are no further thermodynamically relevant contributions.
Again we do not know a general answer. However, there are topological
arguments supporting the existence of such additional
contributions in certain higher-dimensional systems, and suggesting
their absence otherwise. Furthermore,
we can perform a quantitative comparison for each specific spin system
separately with numerical results for finite systems.

\begin{table}
\begin{center}
\caption
{The degeneracies and the energy gaps for various finite $s=1/2$ 
frustrated bilayer lattices:
exact diagonalization data for finite systems
with $J_2 = 5 \,J_1$ vs hard-square predictions.
$N$ is the number of sites in the spin lattice,
DGS is the degeneracy of the ground state as it follows
from exact diagonalization for a given $S^z$,
\# HSS is the number of configurations with $N/2-S^z$ hard squares,
$\Delta$ is the energy gap between the ground state and the first excited one
in units of $J_1$. The dots for $N=64$ indicate omitted sectors
with $16 \le S^z \le 26$.
\label{tab3}}
\vspace{5mm}
\begin{tabular}[t]{|c||c|c|c|c|} \hline
 $N$ & $S^z$ & DGS  & $\Delta$ & \# HSS \\ \hline \hline 
  16 &   7   &    8 &  1.0     &    8   \\
     &   6   &   12 &  1.0     &    12  \\
     &   5   &    8 &  2.0     &    8   \\
     &   4   &    2 &  3.0     &    2   \\ \hline
  20 &   9   &   10 &  1.0     &    10  \\ 
     &   8   &   25 &  1.0     &    25  \\
     &   7   &   20 &  1.0     &    20  \\        
     &   6   &   10 &  2.0     &    10  \\ 
     &   5   &    2 &  3.0     &     2  \\ \hline     
\end{tabular}
\begin{tabular}[t]{|c||c|c|c|c|} \hline
 $N$ & $S^z$ & DGS  & $\Delta$ & \# HSS \\ \hline \hline 
  32 &  15   &   16 &  1.0     &    16  \\ 
     &  14   &   88 &  1.0     &    88  \\ 
     &  13   &  208 &  1.0     &   208  \\ 
     &  12   &  228 &  1.0     &   228  \\ 
     &  11   &  128 &  1.0     &   128  \\  
     &  10   &      &  1.0     &    56  \\  
     &   9   &      &  2.0     &    16  \\  
     &   8   &      &  3.0     &     2  \\ \hline     
\end{tabular}
\begin{tabular}[t]{|c||c|c|c|c|} \hline
 $N$ & $S^z$ & DGS  & $\Delta$ & \# HSS \\ \hline \hline 
  64 &  31   &   32 &  1.0     &    32  \\ 
     &  30   &  432 &  1.0     &   432  \\ 
     &  29   & 3232 &  1.0     &  3232  \\ 
     &  28   &      &  1.0     & 14840  \\ 
     &  27   &      &          & 43904  \\
     &       &      &          & \dots  \\ \hline
\end{tabular}
\end{center}
\end{table}

%

{}From exact diagonalization data for finite systems with periodic
boundary conditions we infer that
for some of the lattices there are no additional states, or at most one or
two additional states in the subspace $S^z=Ns-1$. A perfect correspondence
is found for the diamond chain and the frustrated two-leg ladder with
$J_2>2J_1$ as well as the sawtooth chain and certain two-dimensional models
with finite localization regions such as
the frustrated bilayer lattice with $J_2>4J_1$
(see Ref.~\cite{lm_14} and Table~\ref{tab3}). In other cases such as
the kagom\'{e}-like chains
there are only one or two extra states which can be traced
to one-magnon states from the dispersive band
which at one or two values of the wave-vector have the same energy
as the states from the flat band. In these cases which include in particular
the one-dimensional models, a quantitatively accurate description is
expected from the smallest-size localized magnons.

\begin{table}
\begin{center}
\caption
{The degeneracies and the energy gaps for the $s=1/2$ kagom\'{e} lattice:
exact diagonalization data for finite systems vs hard-hexagon predictions.
$N$ is the number of sites in the spin lattice,
DGS is the degeneracy of the ground state as it follows from exact
diagonalization for a given $S^z$,
\# HHS is the number of configurations with $N/2-S^z$ hard hexagons,
$\Delta$ is the energy gap between the ground state and the first excited one
in units of $J$.
Dots indicate some sectors with $S^z \ge 7\,N/18$ which have been
omitted for larger values of $N$.
For the two-magnon sector one has \# HHS$\,=N^2/18 -7\,N/6$ and 
DGS$\,=N^2/18 -N/2+1$. 
\label{tab1}}
\vspace{5mm}
\begin{tabular}[t]{|c||c|c|c|c|} \hline
 $N$ & $S^z$ & DGS  & $\Delta$ & \# HHS  \\ \hline \hline 
  36 &  17   &   13 & 0.500    &   12    \\ 
     &  16   &   55 & 0.182    &   30    \\ 
     &  15   &   71 & 0.055    &   16    \\ 
     &  14   &    8 & 0.034    &    3    \\ \hline
  45 &  43/2 &   16 & 0.251    &   15    \\ 
     &  41/2 &   91 & 0.123    &   60    \\ 
     &  39/2 &  201 & 0.035    &   60    \\ 
     &  37/2 &  110 & 0.011    &   15    \\
     &  35/2 &    4 & 0.012    &    3    \\ \hline
\end{tabular}
\begin{tabular}[t]{|c||c|c|c|c|} \hline
 $N$ & $S^z$ & DGS  & $\Delta$ & \# HHS  \\ \hline \hline 
  54 &  26   &   19 & 0.177    &   18    \\ 
     &  25   &  136 & 0.091    &   99    \\ 
     &  24   &  430 & 0.025    &  180    \\
     &  23   &  513 & 0.009    &   99    \\
     &  22   &  119 & 0.003    &   18    \\
     &  21   &    4 & 0.012    &    3    \\ \hline
  63 &  61/2 &   22 & 0.297    &   21    \\ 
     &  59/2 &  190 & 0.128    &  147    \\
     &  57/2 &  785 & 0.050    &  406    \\
     &       &      &          &  \dots  \\ \hline
\end{tabular}
\begin{tabular}[t]{|c||c|c|c|c|} \hline
 $N$ & $S^z$ & DGS  & $\Delta$ & \# HHS  \\ \hline \hline 
 108 &  53   &   37 & 0.177    &   36    \\ 
     &  52   &  595 & 0.095    &  522    \\
     &       &      &          & \ldots  \\ \hline
 192 &  95   &   65 & 0.101    &   64    \\ 
     &  94   & 1953 & 0.067    & 1824    \\
     &       &      &          & \ldots  \\ \hline
\end{tabular}
\end{center}
\end{table}

%
\begin{table}
\begin{center}
\caption
{The degeneracies and the energy gaps for various finite $s=1/2$
checkerboard lattices:
exact diagonalization data for finite systems vs large-hard-square predictions.
$N$ is the number of sites in the spin lattice,
DGS is the degeneracy of the ground state as it
follows from exact diagonalization for a given $S^z$,
\# LHSS is the number of configurations with $N/2-S^z$ large-hard-squares,
$\Delta$ is the energy gap between the ground state and the first excited one
in units of $J$.
Dots indicate some sectors with $S^z \ge 3\,N/8$ which have been
omitted for larger values of $N$.
For the two-magnon sector one has \# LHSS$\,=N^2/8 -9\,N/4$ and 
DGS$\,=N^2/8 -5\,N/4+1$. 
\label{tab2}}
\vspace{5mm}
\begin{tabular}[t]{|c||c|c|c|c|} \hline
 $N$ & $S^z$ &  DGS & $\Delta$ & \# LHSS \\ \hline \hline
  40 &  19   &   21 &  0.882   &  20      \\
     &  18   &  151 &  0.222   &  110     \\
     &  17   &  411 &  0.071   &  180     \\
     &  16   &  246 &  0.014   &  85      \\
     &  15   &    4 &  0.028   &  4       \\ \hline
\end{tabular}
\begin{tabular}[t]{|c||c|c|c|c|} \hline
 $N$ & $S^z$ &  DGS & $\Delta$ & \# LHSS \\ \hline \hline
  64 &  31   &   33 &  0.586   &  32      \\
     &  30   &  433 &  0.176   &  368     \\
     &  29   & 2833 &  0.082   &  1888    \\
     &  28   & 9273 &  0.026   &  4392    \\
     &  27   &      &          &  4224    \\
     &  26   &      &          &  1520    \\
     &  25   &      &          &  224     \\
     &  24   &      &          &  12      \\ \hline
\end{tabular}
\begin{tabular}[t]{|c||c|c|c|c|} \hline
 $N$ & $S^z$ &  DGS & $\Delta$ & \# LHSS \\ \hline \hline
 144 &  71   &   73 &  0.268   &  72      \\
     &  70   & 2413 &  0.128   &  2268    \\
     &  69   &      &          &  41208  \\
     &       &      &          &  \ldots  \\ \hline
 256 & 127   &  129 &  0.152   &  128     \\
     & 126   & 7873 &  0.098   &  7616    \\
     & 125   &      &          &  279936  \\ 
     &       &      &          &  \ldots  \\ \hline
\end{tabular}
\end{center}
\end{table}

%

In contrast, the topology of certain two-dimensional lattices
permits the construction of additional, nested localized-magnon states
\cite{lm_07,rev_zhito}.
An examples of such a `defect' state is shown
in Fig.\ \ref{fig03}b.
Such defect states occur first in the two-magnon sector and,
once again, the
issue of linear independence from the smallest-area localized magnon
states arises.
We would like to point out that boundary conditions again play a crucial
role: for open boundary conditions, the two-magnon state
obtained from the
two nested loops at the lower left corner of Fig.\ \ref{fig03}b is
indeed a new state. However, for the periodic boundary conditions which
we are considering here, such states
can be expressed as a linear combination of smallest-area localized
two-magnon states. Let us explain this in the case of the checkerboard
lattice. First note that in a finite checkerboard lattice with periodic
boundary conditions, a magnon state localized on a closed loop $L$
such as the large loop in Fig.\ \ref{fig03}b
can be written as a linear combination of all
magnon states localized on the smallest squares contained in $L$. But this
linear combination is not unique since the set of all smallest
localized one-magnon states is not linearly independent. Another linear combination
yielding the same state could run over all magnon states $|q\rangle$
localized on the smallest squares $q$ {\it not} contained in $L$.
Consider in particular the octagonal loop $L$ containing a smallest
square $q_0$ with a magnon state $|q_0\rangle$ in its center shown in
Fig.~\ref{fig03}b. Then the state of the two magnons localized on $L$ and
$q_0$ can be written as a linear combination of large-hard-square
two-magnon states $|q\rangle\,|q_0\rangle$ where $|q\rangle$ runs
over all one-magnon states of smallest area not contained in $L$.

Note that the derivation of the linear relation
relies on the absence
of obstacles such as open boundaries or other localized magnons outside
the two-magnon defect state. We therefore believe that the majority
of many-magnon states constructed with these defect states
(in particular the
three-magnon state of Fig.~\ref{fig03}b)
is linearly independent of many-magnon states
containing only those of smallest area.
The many-defect states are therefore expected to
yield another finite (even if small) contribution to the ground-state
entropy at the saturation field. Indeed, numerical data for finite lattices
exhibits a larger ground-state degeneracy than predicted from the
effective hard-object description
(see Table~\ref{tab1} for the kagom\'e lattice and Table~\ref{tab2} for
the checkerboard lattice).
This difference should remain relevant in the thermodynamic limit according to
the preceding argument based on the defect states.

Before we proceed, we would like to comment on some specific sectors for
the kagom\'e lattice (Table~\ref{tab1}) and the checkerboard lattice
(Table~\ref{tab2}). Firstly, the ground-state
degeneracy (DGS) in the one-magnon sector is $N/3+1$ ($N/2+1$)
for the kagom\'e (checkerboard) lattice. There are two possible
interpretations (see also \cite{rev_zhito}): in momentum space this
corresponds to the $N/3$ ($N/2$) states of the flat branch plus
one additional state where the next dispersive branch touches
the flat branch; in real space this corresponds to
$N/3$ ($N/2$) smallest localized-magnon states subject
to one linear relation plus two additional states which wind once around
the boundaries.
Secondly, the exact degeneracy in the two-magnon sector is given by 
$N^2/18 -N/2+1$ for the kagom\'e lattice and $N^2/8 -5\,N/4+1$ for the 
checkerboard lattice. The difference with respect to the number of 
smallest-area localized-magnon configurations is again due to states 
with non-trivial winding. Let us explain this briefly for the 
checkerboard lattice. First, there are $N$  configurations of the form 
$|q\rangle|w_x\rangle$ or $|q\rangle|w_y\rangle$ where $q$ runs over all 
squares and $w_\alpha$ is an arbitrary
loop which winds around boundary $\alpha=x$, $y$ sufficiently far away 
from $q$. As we will explain in detail elsewhere, taking into account 
states $|w_\alpha \rangle |w'_\alpha \rangle$ with double winding
together with one special diagonal state
and the linear relations between these states, one finds $N+1$ additional
linearly independent
two-magnon states with non-trivial winding, {\it i.e.}, exactly the same 
number as observed numerically. The difference $2\,N/3+1$ between the 
number of configurations of two hard hexagons and the exact ground-state 
degeneracy in the two-magnon sector for the kagom\'e lattice 
can be explained in an analogous way.
Finally, let us look at the sector with the closest packing of
localized magnons, {\it i.e.}, $S^z = 7\,N/18$ for the kagom\'e lattice
and $S^z = 3\,N/8$ for the checkerboard lattice.
For the $N=40$ checkerboard lattice we find the same number of ground states
and large-hard-square states for $S^z=15$, as expected for a closest
packing. By sharp contrast, the kagom\'e lattice gives rise to
8 ground states for $N=36$, and 4 for $N=45$, $54$
in the sector with $S^z = 7\,N/18$ while one expects only 3 for
the magnon crystal \cite{lm_02,rev_rich_etal,rev_zhito}.
Since the $N=36$, $45$ and $54$ kagom\'e lattices should be sufficiently
large to eliminate boundary artifacts, the origin of the additional state(s)
is unclear at present. In particular, it remains to be clarified whether
(essentially) all ground states are described by localized magnons
if all topological non-trivial configurations (including defect
states and states with non-trivial winding) are properly accounted for.

One further issue 
is whether the ground-state manifold
is separated from other states by a finite energy gap.
We can again draw some conclusions concerning the energy gap from exact
diagonalization data.
In previous papers \cite{lm_14,lm_16} we introduced 
a measure
for the thermodynamically relevant energy separation $\Delta_{\rm{DOS}}$
between the ground-state manifold and the other eigenstates of the system.
For simplicity, in the present paper we report the energy gap $\Delta$
between the ground-state energy and the next smallest energy level
in each sector $S^z$.
Tables~\ref{tab3}, \ref{tab1}, and \ref{tab2}
present the values of $\Delta$ for some finite frustrated bilayer,
kagom\'{e}, and checkerboard lattices, respectively.
For the frustrated bilayer lattice with $J_2 = 5\,J_1$, the localized
magnons are separated from all higher excitations by an energy gap
$\Delta \ge J_1$ (see Table~\ref{tab3}).
For the kagom\'e lattice one has the additional complication that there
is no gap to the next branch of excitations in the one-magnon sector.
Nevertheless, two-magnon scattering states were estimated to have an
energy gap $\Delta \approx 0.24\,J$ \cite{lm_07,rev_zhito}. While this may
be a valid estimate in the two-magnon sector of the kagom\'e lattice,
in higher sectors there are definitely excitations at substantially lower
energies (see Table~\ref{tab1}). In fact, analysis of further excited
states (not shown here) indicates the onset of a thermodynamically
relevant density of states at energies of the order of only $10^{-2}\,J$.

\section{Low-temperature strong-field thermodynamics. Lattice-gas description}

\label{sec:LT}

As mentioned above,
the smallest-area localized-magnon states may dominate the low-temperature
thermodynamics in the vicinity of the saturation field.
After having checked their linear independence we
would like to discuss their contribution
to the canonical partition function of the spin system in more detail.
We start from  Eq.~(\ref{05}), where
this contribution is given. We emphasize once again that this formula describes
the low-temperature thermodynamics near the saturation field accurately
provided
that (i)~there are no other ground states (apart from
the smallest-area localized-magnon states) in the corresponding sectors of $S^z$
or that the contribution of such extra states is vanishingly small as $N\to\infty$,
and that (ii)~excited states in these sectors
are separated by a finite energy gap from the ground states. 
In Eq.~(\ref{05}) $g_N(n)$ is the degeneracy of $n$ isolated smallest-area
localized magnons on a spin lattice of $N$ sites.
It is useful to consider $g_N(n)$
as the canonical partition function $Z(n,{\cal{N}})$
of $n$ hard-core objects on an auxiliary lattice of ${\cal{N}}\propto N$ sites
(${\cal{N}}=N/3$, $N/2$, $N/3$, and $N/2$ 
for the diamond chain, the frustrated two-leg ladder, the kagom\'{e}-like chain,
and the frustrated bilayer lattice, respectively).
We can write the grand-canonical partition function
of hard-core objects on this lattice as
$\Xi(T,\mu,{\cal{N}})
=\sum_{n=0}^{n_{\max}}g_N(n)\exp\left(\mu n/kT\right)$,
where
$\mu$ is the chemical potential of the hard-core objects.
The simple reason why a hard-core object lattice-gas description emerges here
is the existence of the ``hard-core rules''
which the localized-magnon states must respect
in order to be eigenstates of the spin Hamiltonian.
Note that
these rules differ for various spin lattices: 
the diamond chain of Fig.~\ref{fig01}a gives rise to hard monomers, 
the frustrated ladder (Fig.~\ref{fig01}b) 
and the kagom\'{e}-like chain (Fig.~\ref{fig01}c) are described by hard dimers, 
while the hard-core objects for two-dimensional lattices are illustrated in Fig.~\ref{fig02}.
Using $\Xi(T,\mu,{\cal{N}})$
we arrive at the basic relation between the localized-magnon contribution
to the canonical partition function of the spin model
and the grand-canonical partition function of the corresponding hard-core
object lattice-gas model,
\begin{eqnarray}
\label{06}
Z_{{\rm{lm}}}(T,h,N)
=\exp\left(-\frac{E_{{\rm{FM}}}-hNs}{kT}\right)\,\Xi(T,\mu,{\cal{N}}) \, ,
\end{eqnarray}
with $\mu=h_1-h$.
From Eq.~(\ref{06}) we find the Helmholtz free energy of the spin system
\begin{eqnarray}
\label{07}
\frac{F_{{\rm{lm}}}(T,h,N)}{N}
=\frac{E_{{\rm{FM}}}}{N}-hs-kT \, \frac{{\cal{N}}}{N} \,
 \frac{\ln\Xi(T,\mu,{\cal{N}})}{{\cal{N}}}.
\end{eqnarray}
The entropy $S$, the specific heat $C$,
the magnetization $M=\langle S^z\rangle$,
and the susceptibility $\chi$ follow from (\ref{07})
according to usual formulae,
$S_{{\rm{lm}}}(T,h,N)=-\partial F_{{\rm{lm}}}(T,h,N)/\partial T$,
$C_{{\rm{lm}}}(T,h,N)=T\partial S_{{\rm{lm}}}(T,h,N)/\partial T$,
$M_{{\rm{lm}}}(T,h,N)=Ns-kT\partial \ln\Xi(T,\mu,{\cal{N}})/\partial\mu$,
and
$\chi_{{\rm{lm}}}(T,h,N)=\partial M_{{\rm{lm}}}(T,h,N)/\partial h$,
respectively.
Note that the thermodynamic quantities of the spin system 
depend on the temperature $T$ and the magnetic field $h$
essentially
only through the combination $x=(h_1-h)/kT$ \cite{lm_14}.

To test the hard-object description, we have performed
full diagonalization of spin-1/2 isotropic Heisenberg systems ({\it i.e.},
$\Delta=1$ in Eq.~(\ref{01})), imposing periodic boundary conditions.
For the diamond chain and the frustrated ladder we have also
exploited the local conservation laws. First we rewrite the Hamiltonian
in terms of the total spin on the vertical dimers in Fig.~\ref{fig01}
\cite{HMT00}. For a local spin 1/2, each dimer can only be in the singlet
or triplet state. A singlet cuts the system into smaller fragments.
It is therefore sufficient to compute the spectra of one periodic fragment
where all vertical dimers are in the triplet state, and smaller open
fragments where all consecutive dimers are again in the triplet state.
In this manner it requires only a moderate effort
to obtain the complete spectra for a diamond chain
and a frustrated ladder with $N=24$, while
$N=24$ would be inaccessible for the diamond chain with a full
diagonalization of the original model.

\subsection{Hard-monomer universality class}

First we consider the hard-monomer universality class which includes
the diamond chain, the dimer-plaquette chain and the
square-kagom\'{e} lattice. The hard-monomer restriction
for these lattices means
that it is forbidden to have two (or more) localized magnons
in the same trap.
We focus on the diamond chain (see Fig.~\ref{fig01}a), and refer
the interested reader to Ref.~\cite{lm_14} for the other lattices.
A straightforward calculation yields
the grand-canonical partition function of a gas of hard monomers
\begin{eqnarray}
\label{08}
\Xi(T,\mu,{\cal{N}})
=\left(1+\exp\frac{\mu}{kT}\right)^{{\cal{N}}}.
\end{eqnarray}
Explicit analytic expressions for thermodynamic quantities can be obtained
easily \cite{lm_14}. Note that
the thermodynamic quantities per site are independent of the system size $N$.
At the saturation field we have a residual ground-state entropy
$S_{{\rm{lm}}}(T,h_1,N)/kN=({\cal{N}}/N)\ln 2$.
The specific heat $C_{{\rm{lm}}}(T,h_1,N)/kN$
exhibits two identical maxima of height $\approx 0.43922884 {\cal{N}}/N$
at $x\approx \pm 2.39935728$.

\begin{figure}
\begin{center}
\includegraphics[clip=on,width=5.3cm,angle=0]{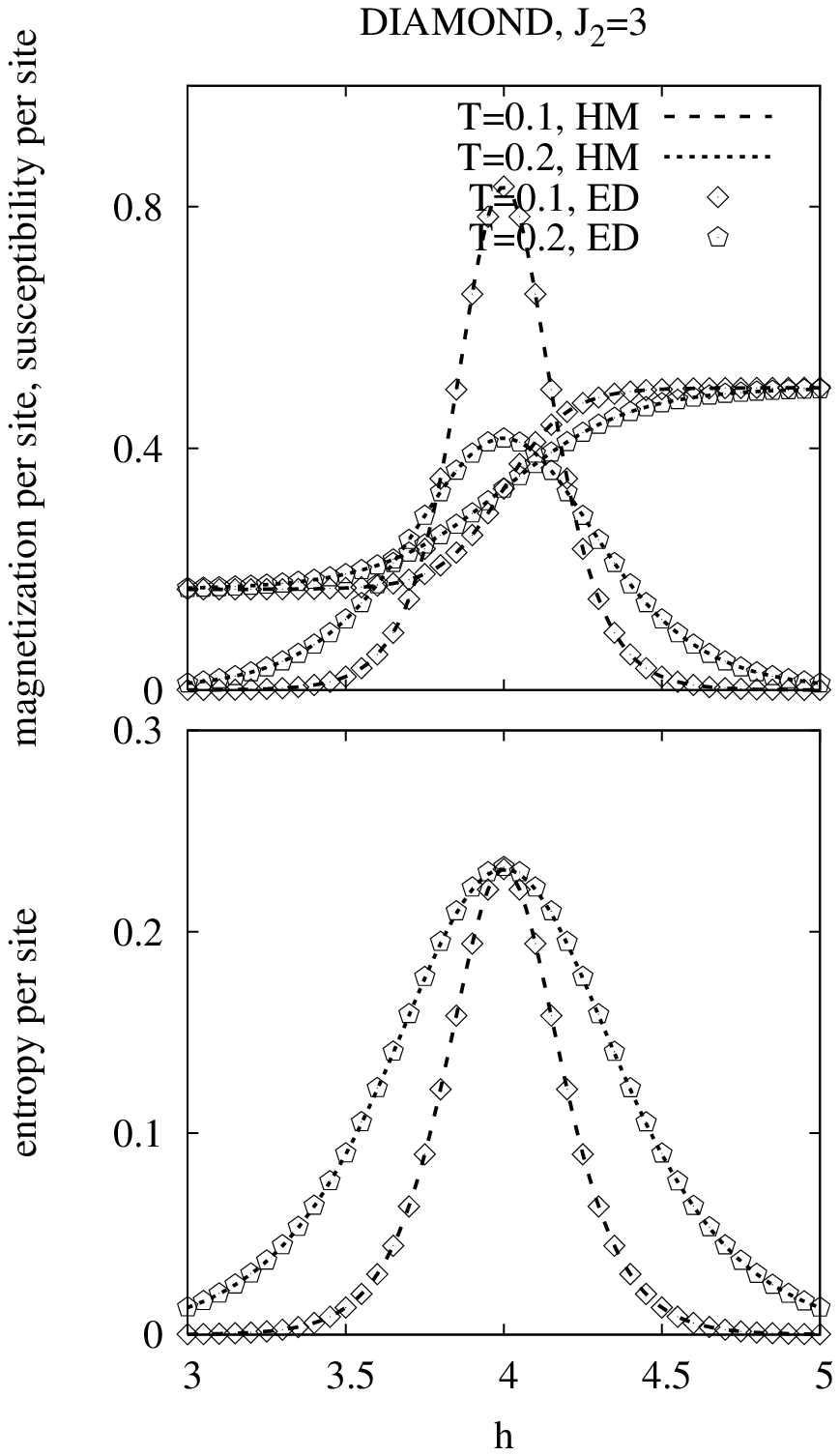}
\includegraphics[clip=on,width=5.3cm,angle=0]{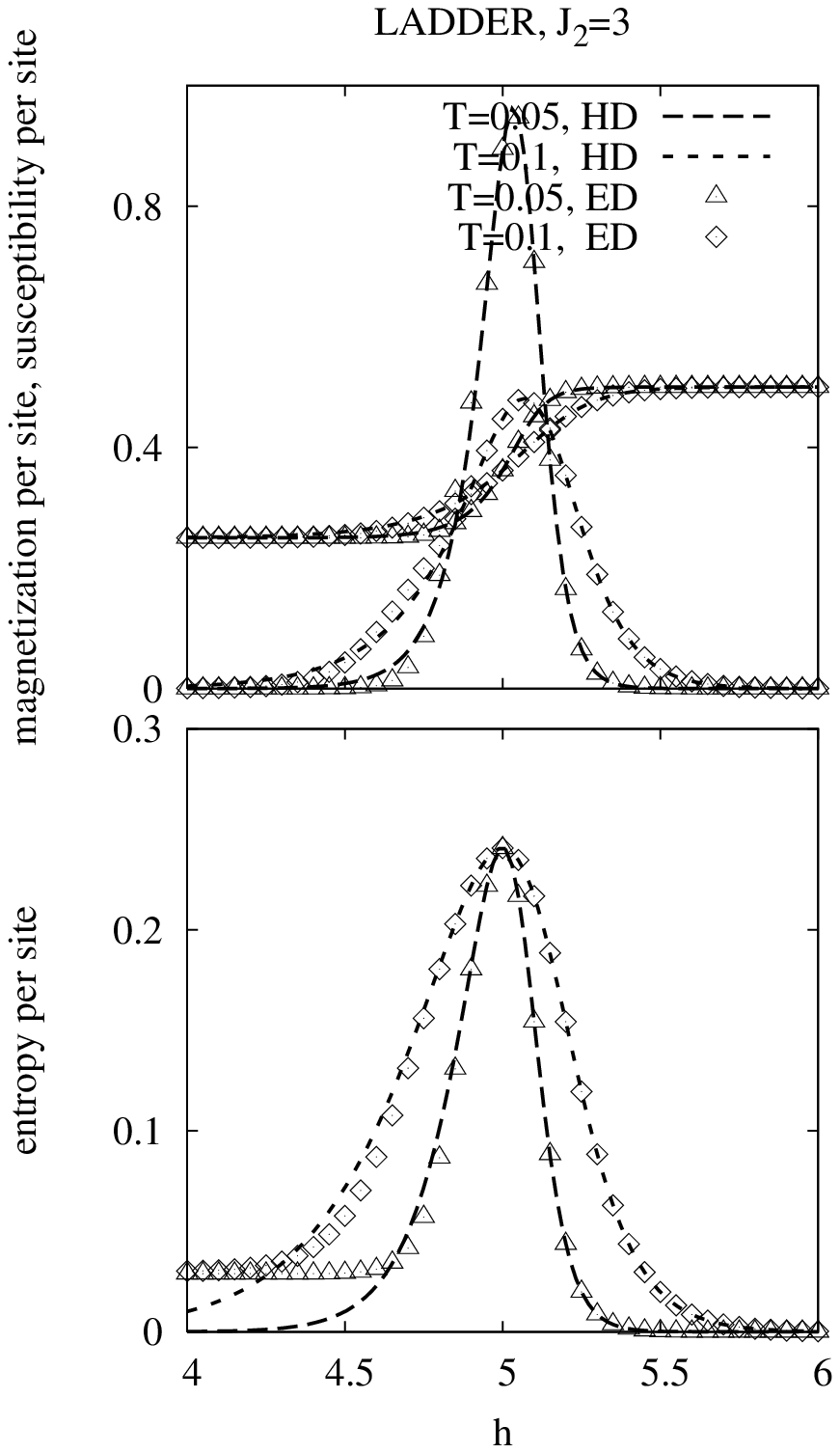}
\includegraphics[clip=on,width=5.3cm,angle=0]{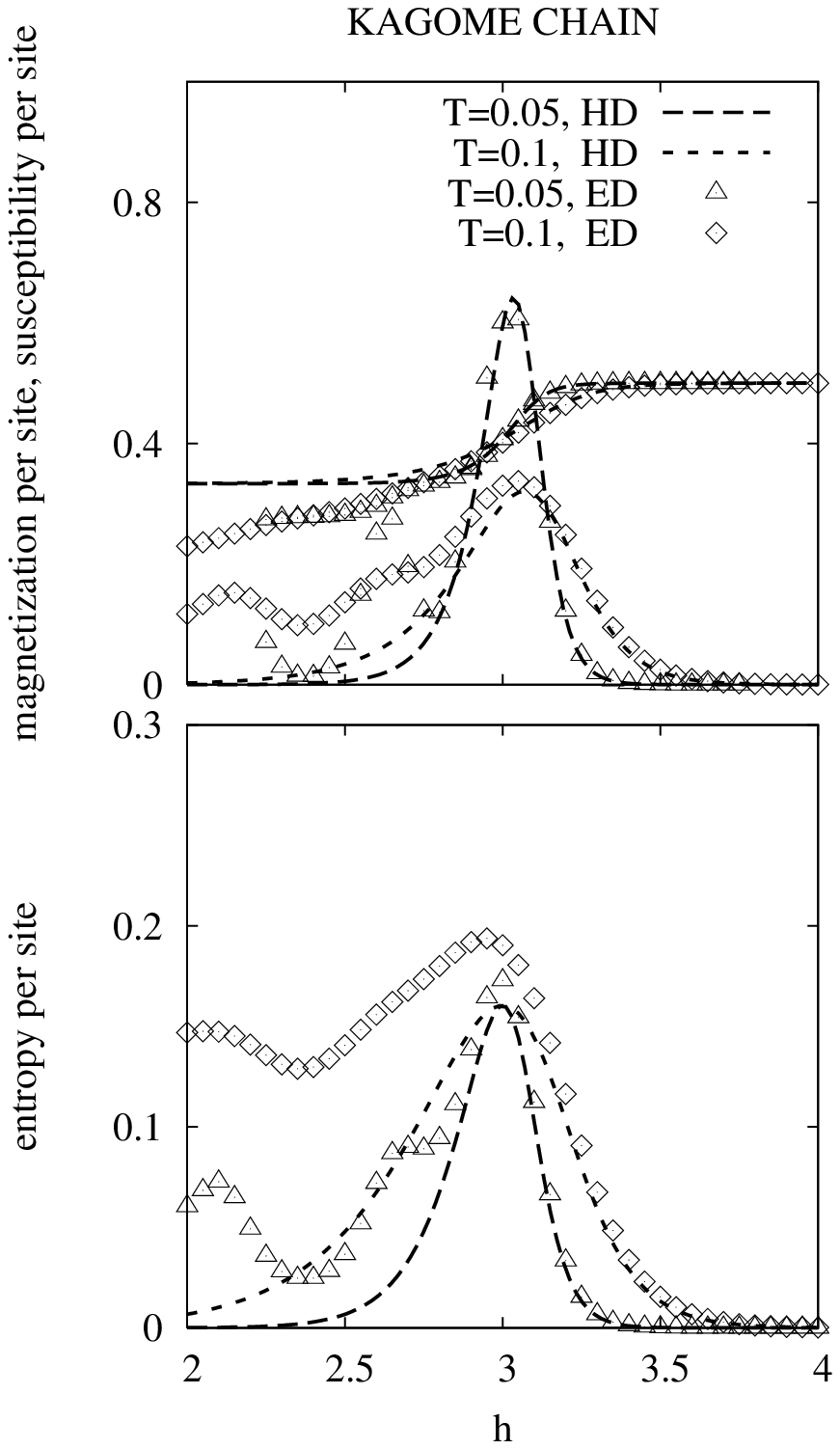}\\
\vspace{5mm}
\includegraphics[clip=on,width=5.3cm,angle=0]{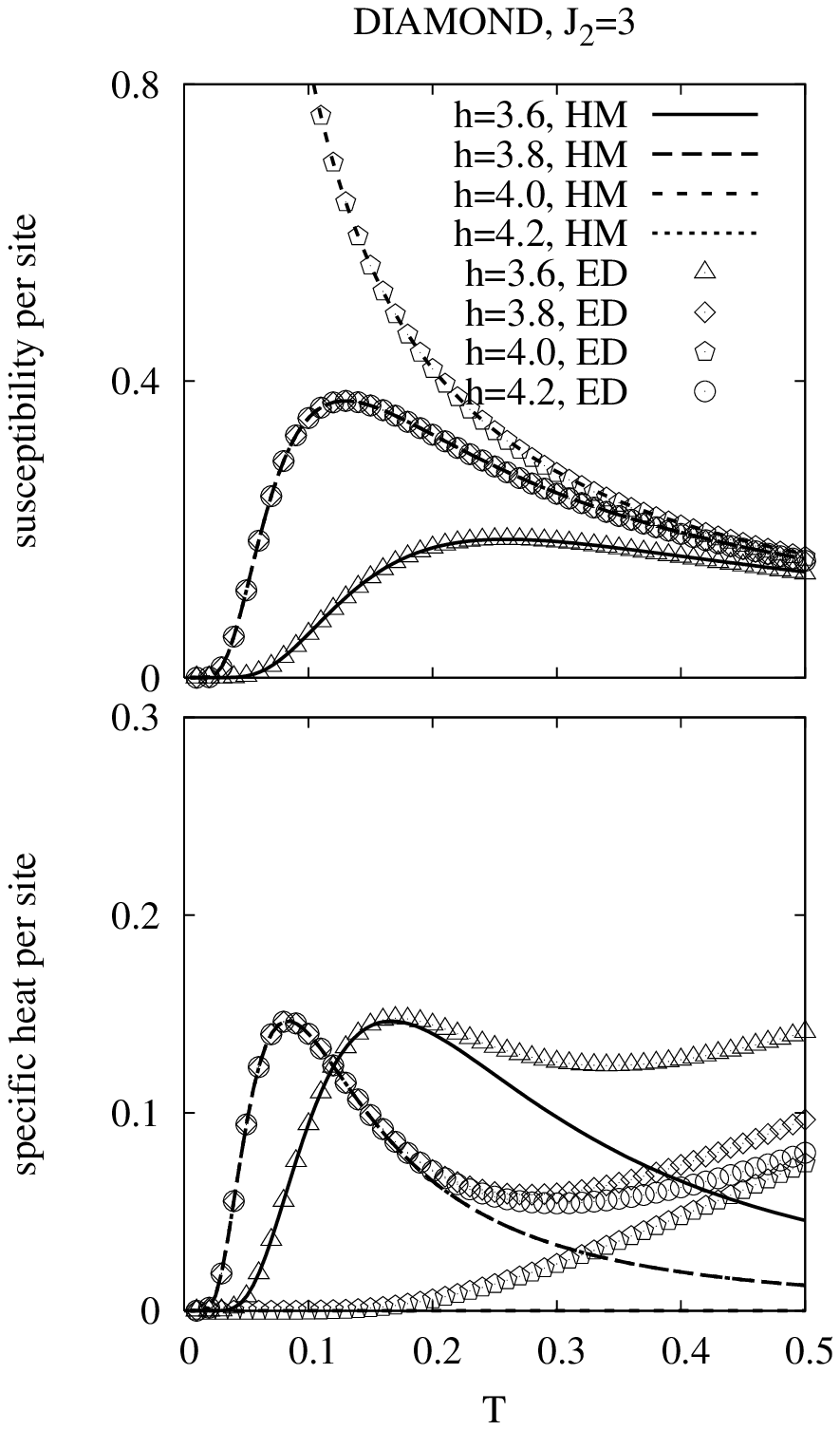}
\includegraphics[clip=on,width=5.3cm,angle=0]{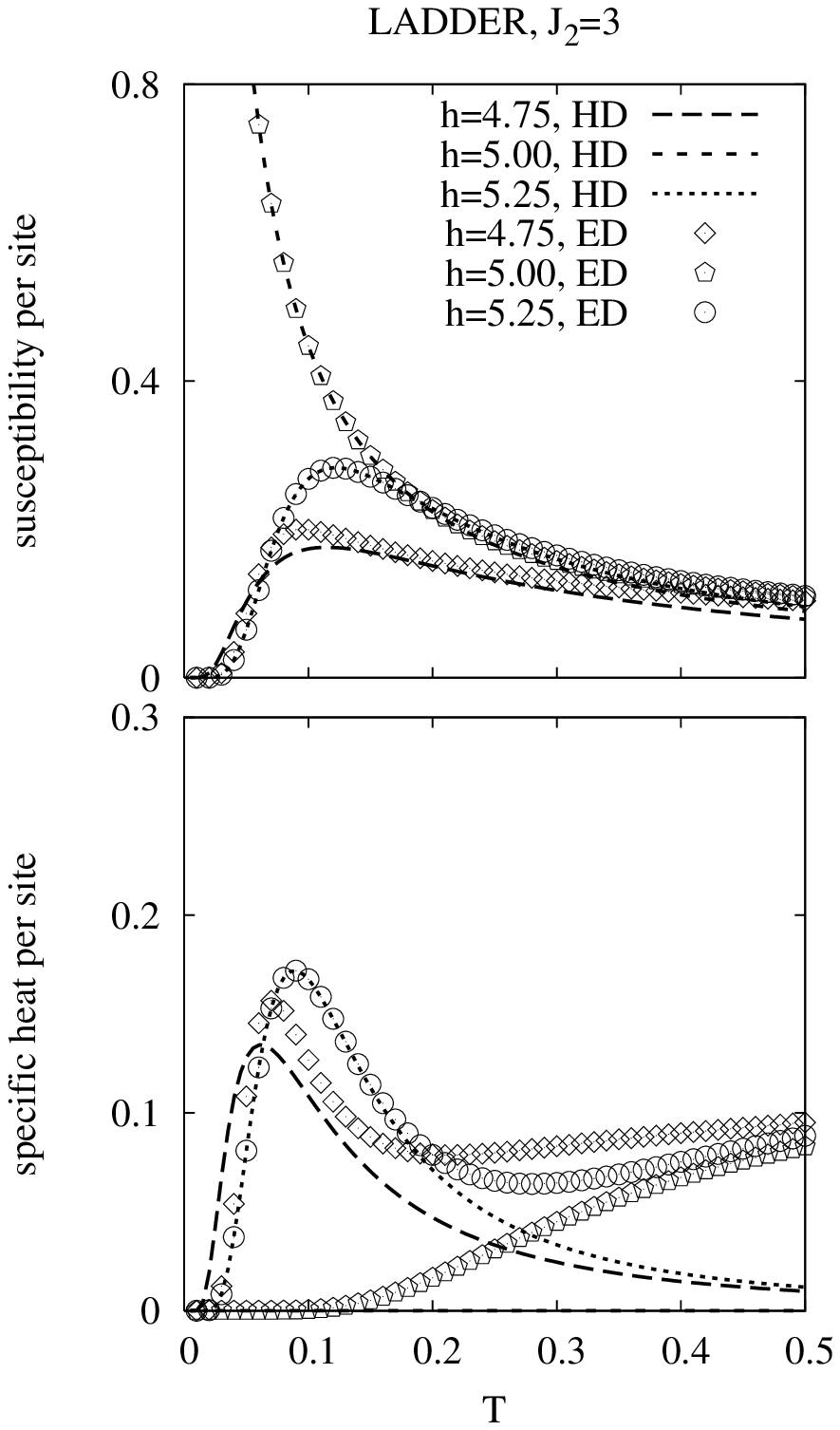}
\includegraphics[clip=on,width=5.3cm,angle=0]{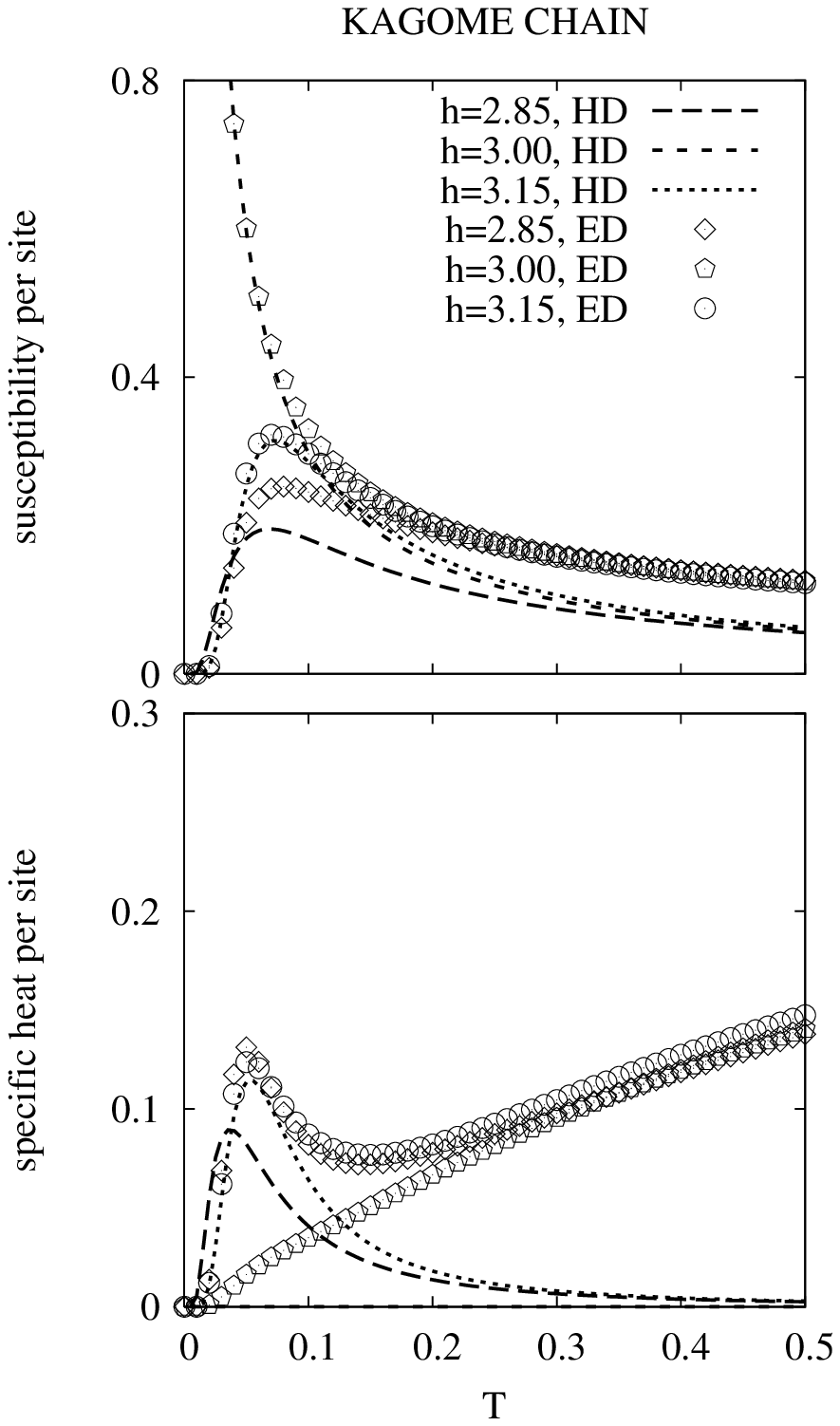}
\caption
{$M(T,h,N)/N$, $\chi(T,h,N)/N$, and
$S(T,h,N)/kN$ vs $h$ at low temperatures;
$\chi(T,h,N)/N$ and
$C(T,h,N)/kN$ vs $kT$ around the saturation field.
{}From left to right:
diamond chain with $J_1=1$, $J_2=3$,
frustrated two-leg ladder with $J_1=1$, $J_2=3$,
kagom\'{e}-like chain with $J=1$.
We set the field range $h_1-1\le h\le h_1+1$
and the temperature range $0\le kT\le 0.5$.
The exact diagonalization (ED) data (symbols) refer to finite systems
of sizes $N=24$ (diamond chain and frustrated ladder) and
$N=18$ (kagom\'{e}-like chain).
The analytical predictions for hard monomers (HM) and one-dimensional
hard dimers (HD) (${\cal{N}}\to\infty$) are shown by lines.}
\label{fig04}
\end{center}
\end{figure}

Some typical dependencies of the thermodynamic quantities on the field and
the temperature
for the diamond chain with $J_1=1$, $J_2=3$ and the corresponding
hard-monomer data obtained on the basis of Eq.~(\ref{08})
are shown in Fig.~\ref{fig04}, left column. Deviations between
exact diagonalization (ED) and hard monomers (HM) are observed only
in the specific heat $C$, and only for $kT \gtrsim 0.2\,J_1$
(lowest panel in the left column of Fig.~\ref{fig04}). Note furthermore
that thermodynamic quantities are symmetric under $x \to -x$ within
the hard-monomer picture. In particular, the hard-monomer description
yields a susceptibility $\chi$ and a specific heat $C$ for the diamond chain
which coincide at $h=3.8\,J_1$ and $h=4.2\,J_1$.

\subsection{One-dimensional hard-dimer universality class}

The frustrated two-leg ladder, the kagom\'{e}-like chains, and
the sawtooth chain
belong to the one-dimensional hard-dimer universality class,
{\it i.e.}, the rules for the localized magnons obey the restrictions
for rigid dimers on a one-dimensional lattice:
each trapping cell can only be occupied by one localized magnon
and neighboring trapping cells cannot be simultaneously occupied by localized
magnons. We focus
on the frustrated two-leg ladder and the kagom\'{e}-like chain of type I
(see Figs.~\ref{fig01}b,~\ref{fig01}c).
The grand-canonical partition function of one-dimensional hard dimers
can be calculated with the help of the transfer-matrix method:
\begin{eqnarray}
\label{09}
\Xi(T,\mu,{\cal{N}})
=\lambda_1^{{\cal{N}}}+\lambda_2^{{\cal{N}}},
\;\;\;
\lambda_{1,2}=\frac{1}{2}\pm\sqrt{\frac{1}{4}+\exp\frac{\mu}{kT}}.
\end{eqnarray}
Explicit analytic expressions for thermodynamic quantities
can again be obtained easily. Note that for hard dimers
the thermodynamic quantities per site depend on the size ${\cal N}$.
In the thermodynamic limit only the largest eigenvalue $\lambda_1$ plays a role.
Explicit expressions in this limit can be found in
Refs.~\cite{lm_06,rev_zhito} for the sawtooth chain
and in Ref.~\cite{lm_14} for the general case.
At the saturation field one finds a residual ground-state entropy
$S_{{\rm{lm}}}(T,h_1,N)/kN=({\cal{N}}/N) \, \ln ((1+\sqrt{5})/2)$.
The specific heat $C_{{\rm{lm}}}(T,h_1,N)/kN$
exhibits two maxima of height
$\approx 0.34394234 \, {\cal{N}}/N$
(at $x\approx -2.81588498$)
and
$\approx 0.26887020 \, {\cal{N}}/N$
(at $x\approx 4.05258891$).
Some typical dependencies of the thermodynamic quantities on
field and temperature
are shown in Fig.~\ref{fig04}, middle and right columns
for the frustrated ladder with $J_1=1$, $J_2=3$
and the kagom\'{e}-like chain with $J=1$, respectively.
Note that the hard-dimer results are not symmetric around the
saturation field, {\it i.e.}, all thermodynamic quantities are different
for $x$ and $-x$.

For the frustrated ladder with $J_2 = 3\,J_1$,
we observe again systematic differences
between exact diagonalization and hard dimers in the specific heat
$C$ at high temperatures (lowest panel in the middle column of Fig.~\ref{fig04}).
The remaining differences for low temperatures and $h < h_1$ are due
to finite-size effects, since in Fig.~\ref{fig04} we compare finite
spin systems with infinite hard-dimer systems. If the comparison is
performed for the same system size \cite{lm_14}, better agreement
can be observed. One important source of finite-size effects is
a two-fold degeneracy of the ground state at $M/N=1/4$.
This ground-state degeneracy is evident
in the finite value of the ED results for the entropy $S$ at the left side
of the second panel in the middle column of Fig.~\ref{fig04}.

In the case of the kagom\'e-like chain
(right column of Fig.~\ref{fig04}) we find good agreement
between exact diagonalization and hard dimers on the high-field
side $h \ge h_1$ and sufficiently low temperatures. However, on the
low-field side $h < h_1$ we observe even stronger deviations than
for the frustrated ladder. Indeed, there are two steps at
$h_2 \approx 2.67\,J$ and $h_3 \approx 2.11\,J$
in the zero-temperature magnetization curve of
the $N=18$ kagom\'e chain with $M/N < 1/3$ (see also Fig.~2 of
Ref.~\cite{lm_17}). Since the region $M/N < 1/3$ cannot
be described with hard dimers, the region with $h \lesssim 2.7\,J$ falls
completely outside the validity of the hard-dimer picture.

\subsection{Two-dimensional lattice gases,
hard-square universality class}

The frustrated bilayer lattice, the kagom\'{e} lattice as well as the
checkerboard lattice are even more interesting,
since the corresponding two-dimensional classical hard-core systems
exhibit a second-order finite-temperature order-disorder phase
transition \cite{baxter,large_squares,squares}.
However, recall from Section~\ref{indep} that the spin model exhibits
extra ground states on the kagom\'{e} and
checkerboard lattices which cannot be described
by hard hexagons or large-hard-squares, respectively.
Moreover, the gap to excited states
is pretty small, see Tables~\ref{tab1} and \ref{tab2}.
Therefore, the description of the kagom\'{e} and
checkerboard lattices in terms of hard-core objects is expected to be
only a qualitative one.

\begin{figure}
\begin{center}
\includegraphics[clip=on,width=8.0cm,angle=0]{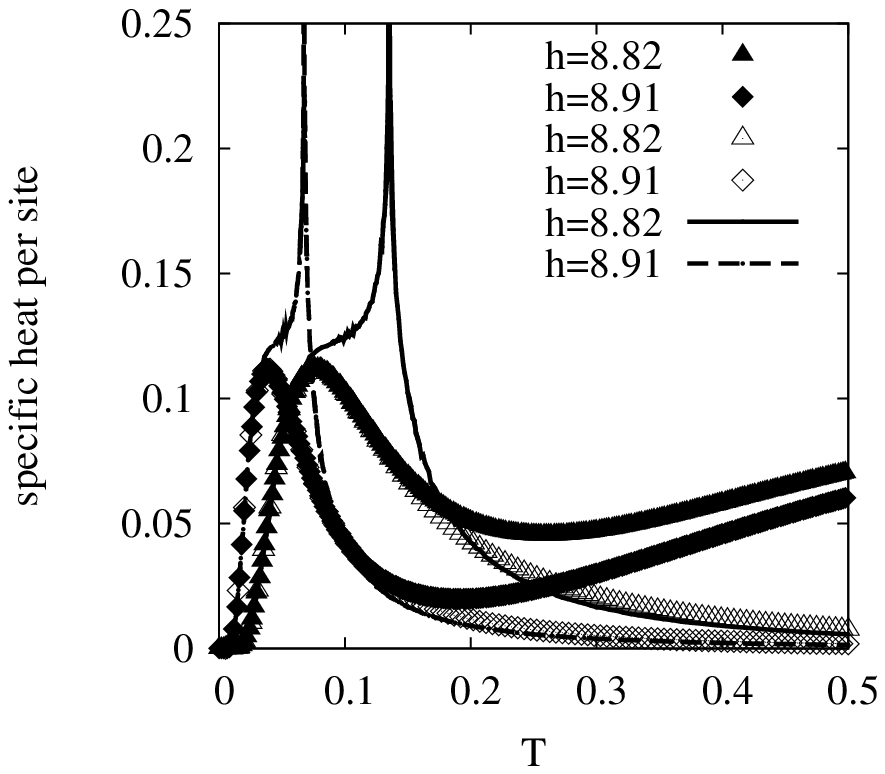}
\includegraphics[clip=on,width=8.0cm,angle=0]{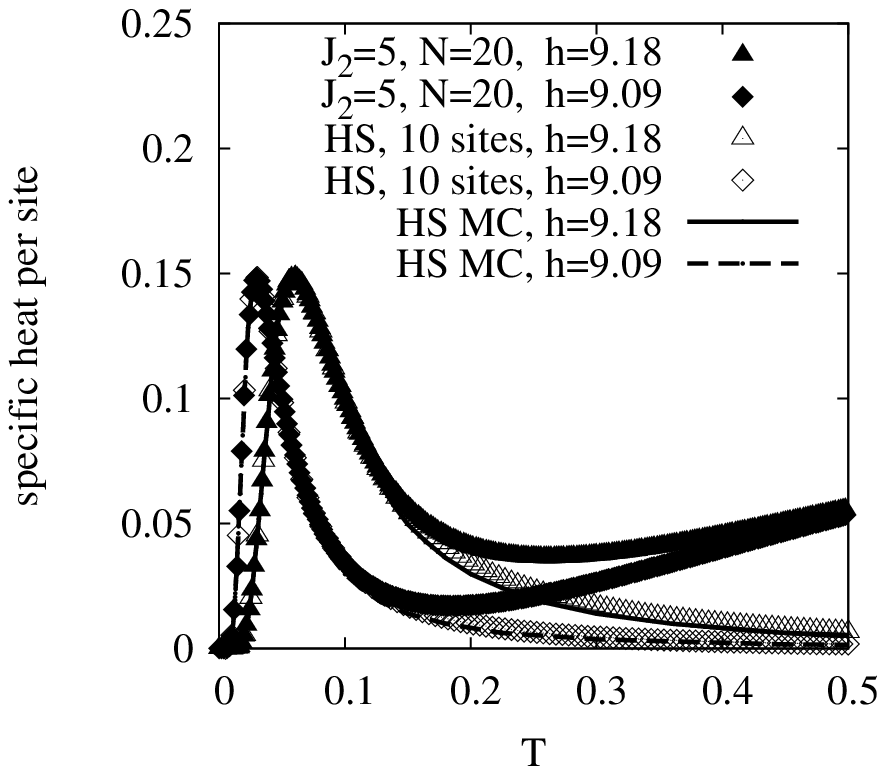}
\caption
{$C(T,h,N)/kN$ vs $kT$ around the saturation field $h_1=9$ for the frustrated
bilayer lattice with $J_1=1$, $J_2=5$.
The exact diagonalization data (filled symbols) refer to a finite spin system
of $N=20$ sites.
The analytical results (empty symbols) refer to a finite hard-square system
of ${\cal{N}}=10$ sites. The Monte Carlo simulation data (lines) are
obtained for the hard-square system on finite lattices
with ${\cal{N}}$ up to $800\times 800$.}
\label{fig05}
\end{center}
\end{figure}

We therefore focus on the frustrated bilayer antiferromagnet as an
example for a finite-temperature order-disorder phase transition \cite{lm_16}.
In this case, exact diagonalization provides clear evidence
that all ground states are mapped onto the hard-square configurations
on a square lattice, see Table~\ref{tab3}. Moreover, the gap to excited
states is large.
Thus, the low-temperature strong-field thermodynamics
should be determined completely by the hard-square problem.
Although we do not know the exact analytical result
for the grand thermodynamical potential
$-kT\ln \Xi(T,\mu,{\cal{N}})/{\cal{N}}$
of hard squares on a square lattice,
the properties of the model are well known \cite{squares,baxter}.
In particular,
the hard-square model exhibits a phase transition at $z_c=\exp(\mu_c/kT)=3.7962\ldots$
between the low-density phase ($z<z_c$),
in which both sublattices of the underlying square lattice are equally occupied,
and the high-density phase ($z>z_c$),
in which one of the sublattices becomes more occupied than the other one.
In spin language,
the phase transition has a purely geometrical origin
and indicates the ordering of localized-magnon states
as their density varies with field or temperature.
The phase transition belongs to the two-dimensional Ising
universality class. Hence, the specific heat should show a
logarithmic singularity at the critical point.
Fig.~\ref{fig05}
shows results for the temperature dependence of the specific heat around
the saturation field for the frustrated bilayer lattice.
The available exact diagonalization data are restricted to rather
small spin systems. However,
they  demonstrate a perfect agreement with the results for the corresponding
finite hard-square model:
for $J_2=5\,J_1$ and $h=0.99\,h_1=8.91\,J_1$ or $h=1.01\,h_1=9.09\,J_1$,
both data sets in Fig.~\ref{fig05} are indistinguishable for temperatures
$kT \lesssim 0.1\, J_1$.
Bigger hard-square systems can be studied using classical
Monte Carlo simulations \cite{taras}.
The left panel of Fig.~\ref{fig05} shows that
a logarithmic singularity in the dependence of $C$ vs $T$ develops
on the low-field side with increasing size of the hard-square model.
Since this transition occurs within the temperature region
where the exact diagonalization data are perfectly reproduced by
the hard-square model,
we expect that this singularity also appears in the spin model in
the thermodynamic limit $N\to\infty$.

To summarize this subsection, 
the frustrated bilayer lattice provides
an example where
a two-dimensional hard-core lattice gas completely covers all low-energy states
of the spin model.
In this model, there is a clear phase transition
which corresponds to a crystallization of hard squares,
{\it i.e.}, magnons localized on the vertical dimers.
We would like to emphasize
that such a phase transition does not contradict
the Mermin-Wagner theorem \cite{mermin} since only a discrete symmetry
is broken spontaneously.
This demonstration of a finite-temperature phase transition in
a two-dimensional interacting many-body spin model
is an interesting example for the impact of the localized-magnon states
on the physical properties of a wide class of frustrated magnets.

\subsection{Region of validity}

Finally, we add some general remarks about the region
of validity of the hard-core lattice-gas description.
Obviously,
this effective picture of the spin model
is accurate only in some region in the $h-T$-plane
around the point $h=h_1$, $T=0$.
The field $h<h_1$,
until which the hard-core object picture should work at $T=0$,
is related to the width of the plateau $h_1-h_2$ preceding the jump in the
ground-state magnetization curve \cite{lm_02,lm_04,lm_05,lm_17}
with $h_2$ being the difference
between the ground-state energy in the sectors $S^z=Ns-n_{\max}$ and $S^z=Ns-n_{\max}-1$.
We could also try  to estimate a characteristic temperature $T^*(h)$,
below which the hard-core object picture should work
at a certain magnetic field $h$.
At $h=h_1$ we can find $T^*$
from the temperature dependence of the specific heat $C$.
Indeed,
the hard-core object prediction is $C=0$ for all temperatures at $h=h_1$.
However, as can be seen in the corresponding panels in Fig.~\ref{fig04}
this is a valid approximation for $kT \lesssim 0.2J_1$ ($kT \lesssim 0.15J_1$) for the
diamond chain (frustrated ladder), whereas the corresponding
temperature region is much smaller for the kagom\'{e}-like chain.
For $h>h_1$ the temperature $T^*$
depends on the difference  $h-h_1$.
Numerical data for finite systems indicates
that $T^*$ increases with growing $h-h_1$.

Lastly,
we note that a description of the low-energy degrees of the
spin systems can be extended
by relaxing the hard-core rules, e.g.\ by
rendering the infinite nearest-neighbor repulsion finite
or permitting double occupation of the auxiliary lattice sites \cite{rev_zhito,lm_14}.
In such a case
we may achieve a better agreement with exact diagonalization data in a wider range of parameters
around the point $h=h_1$, $T=0$,
loosing, however, the universal dependence on $h$ and $T$ only via the
parameter $x=(h_1-h)/kT$.

\section{Conclusions}

\label{sec:Concl}

In this paper, we have discussed the universal properties of some highly
frustrated quantum Heisenberg antiferromagnets supporting localized-magnon
eigenstates. Universal behavior emerges owing to the localized-magnon
states which become the ground states around the saturation field and can be 
separated from the higher-energy states by an energy gap. We find several
universality classes depending on the specific lattice-gas model of
hard-core objects which describes the low-energy degrees of freedom of the
spin model in strong magnetic fields. For the one-dimensional models, the
lattice gas yields a quantitative description of the thermodynamics of the
full spin model close to the saturation field and at sufficiently low
temperatures.

Higher dimensions may be even more interesting since they allow for a
finite-temperature crystallization phase transition of the hard-core
objects. As a two-dimensional example, we have focused on the hard-square
universality class which contains, e.g., the frustrated bilayer quantum
Heisenberg antiferromagnet in the region with sufficiently strong
interlayer coupling \cite{lm_16}. Such a spin model exhibits an
order-disorder phase transition of a purely geometrical origin which
reflects the geometrical ordering of the localized magnons. It turns out
that the phase transition of hard squares belongs to the two-dimensional
Ising universality class and is characterized by a logarithmic singularity
of the specific heat just below the saturation field.

New numerical results and complementary arguments presented in this paper
indicate that the situation is more complicated in the case of the
kagom\'e and checkerboard lattices. Here, there seem to be
thermodynamically relevant contributions to the ground-state manifold
beyond that of hard hexagons and large-hard squares, respectively.
Accordingly, in these cases the gas of smallest hard objects cannot be
expected to yield a quantitatively accurate description of the spin model
in any parameter regime. Still, the universality class of a possible
crystallization phase transition should be predicted correctly by such an
effective low-energy theory.

We believe that the properties of localized magnons elucidate the physics
of frustrated quantum antiferromagnets in high magnetic fields and thus
are useful for a general understanding of related compounds. Even more,
recent experiments on the spin-1/2 (distorted) diamond-chain compound
azurite Cu$_{\rm{3}}$(CO$_{\rm{3}}$)$_{\rm{2}}$(OH)$_{\rm{2}}$
\cite{kikuchi_old,kikuchi} and the frustrated quasi-two-dimensional
spin-$1/2$ antiferromagnet Cs$_{\rm{2}}$CuCl$_{\rm{4}}$ \cite{radu} raise
hopes for a direct comparison with the theoretical models discussed in
this paper, although so far there is no clear experimental observation of
the pronounced quantum effects predicted for low-dimensional spin-$1/2$
antiferromagnets with localized magnons yet.

The Cu$^{2+}$ ions in azurite
Cu$_{\rm{3}}$(CO$_{\rm{3}}$)$_{\rm{2}}$(OH)$_{\rm{2}}$ form infinite
chains with the structure of a spin-$1/2$ distorted diamond chain
\cite{kikuchi_old,kikuchi}. The high-field magnetization $M(h)$ of azurite
exhibits a plateau at 1/3 of the saturation magnetization and a further
steep increase as the magnetic field tends to the saturation value of
about 32.5~T \cite{kikuchi_old,kikuchi}. Note that an experimentally
accessible saturation field is an attractive feature of this compound.
Fits of the magnetization curve and thermodynamic properties with
high-temperature series \cite{HoL01} and numerical results yield the
following estimates for the exchange interactions \cite{kikuchi}:
$J_1^{w}=8.6$~K, $J_1^{s}=19$~K, $J_2=24$~K (in Fig.~\ref{fig01}a the
bonds with weaker interaction $J_1^{w}$ run from south-west to north-east
whereas the bonds with stronger interaction $J_1^{s}$ run from north-west
to south-east). Deviations from the ideal diamond chain geometry are not
necessarily a major problem (see below). However, the quoted estimates for the
exchange interactions are not in the region $J_2 \ge 2\,J_1$ which
is required to render the localized magnons low-energy excitations
(compare also the phase diagram of the distorted diamond chain
\cite{OTTK}).
Nevertheless, the values of the exchange couplings are still under debate
\cite{kikuchi}.

Cs$_{\rm{2}}$CuCl$_{\rm{4}}$ is a frustrated quasi-two-dimensional
spin-1/2 antiferromagnet with a low saturation field of only about
8.5~T \cite{coldea02}. For this compound, measurements of the
low-temperature behavior of the specific heat around the saturation
field have already been performed, exhibiting a strong dependence on the
magnetic field \cite{radu}. The dominant exchange
interactions in Cs$_{\rm{2}}$CuCl$_{\rm{4}}$ correspond to an
anisotropic triangular lattice with strong interactions along one
`chain' direction. For such a model one can construct magnons localized
on the strongly coupled chains, much in the same way as for the frustrated square
lattice \cite{lm_02,rev_rich_etal}. Due to the quantization
of momenta transverse to the chain direction, these localized magnon
states turn out to be high-field ground states for up to at least six coupled
chains in the parameter regime relevant to Cs$_{\rm{2}}$CuCl$_{\rm{4}}$
\cite{coldea02}. The localized magnons cease to be ground states
as one approaches the thermodynamic limit, but they remain low-energy
excitations. However, even in a case such as the frustrated square lattice
where magnons localized on lines are high-field ground states for all
finite systems, these are not thermodynamically relevant, but instead
magnetic order occurs below the saturation field \cite{JaZh04}.
Indeed, Cs$_{\rm{2}}$CuCl$_{\rm{4}}$ exhibits a finite-temperature
magnetic ordering transition below the
saturation field \cite{coldea02} such that inter-plane coupling will
have to be taken into account for a quantitative description
of the low-temperature specific heat \cite{radu}. Nevertheless,
the existence of localized magnons and the strong field-dependence of the
specific heat of Cs$_{\rm{2}}$CuCl$_{\rm{4}}$ \cite{radu} are both
related to the strong frustration of the anisotropic triangular lattice.

With respect to experiments such as those on azurite
it is desirable to be not restricted to the ``ideal geometry''
allowing  existence of exact localized-magnon ground states
and to examine the ``stability'' of our results against small deviations
from the relation which we have imposed on the exchange interactions.
Numerical studies
for finite spin systems \cite{lm_08}
suggest that the main
features coming from the localized-magnon states survive in case of
small deviations
from the ideal lattice geometry.
In general, we can argue
that due to small deviations from the ideal geometry,
the flat magnon band becomes slightly dispersive but the
hard-core constraint is preserved.
Apparently, in this case we are faced with a {\it quantum} hard-core object model
(e.g., the quantum hard-square model studied in Ref.~\cite{henley},
see also references therein).
A study of the corresponding low-energy theories
is beyond the scope of the present paper.

\section*{Acknowledgments}

O.~D. and J.~R. acknowledge the kind hospitality of the MPIPKS-Dresden in the end of 2006.
O.~D. is indebted to Magdeburg University and to Wroclaw University
for their hospitality in the autumn of 2006.
A part of the numerical calculations was performed  
using J\"{o}rg Schulenburg's {\it spinpack}.
We would like to thank M.~E.~Zhitomirsky for useful discussions and
comments on this manuscript.
\\

{\bf Note added:} 
Two closely related preprints \cite{ZhiTsu06,MFAL06}
were submitted
shortly after the present one \cite{our_preprint}. 
Ref.~\cite{ZhiTsu06} confirms
some of the new conclusions of our Section \ref{indep}. Furthermore,
Refs.~\cite{ZhiTsu06,MFAL06} both contain Monte-Carlo results for
the phase transition in the large-hard-square lattice gas. They
agree on the location of the critical point while the universality
class of the phase transition remains under debate.



\begin{thebibliography}{999}
\bibitem{footnote1}
The diamond chain, 
the frustrated ladder and the frustrated bilayer lattice fall into a class of models with
local conservation laws. These local conservation laws can be exploited
in the analysis of such models \cite{gelfand91,HMT00,BBL02,CB02,ToMi02,TKS96,HoL01,dim_pla}.
\bibitem{taras}
The classical Monte Carlo data are taken from Ref.~\cite{lm_16}.
The numerical calculations were performed by T.~Krokhmalskii.
\end{thebibliography}
\end{document}